\documentclass[10pt,twocolumn,superscriptaddress,longbibliography,prx]{revtex4-2}

\usepackage{amsmath, bm, amsfonts, amssymb}     
\usepackage{graphicx}                       
\usepackage{xfrac}                          
\usepackage{grffile}                        
\usepackage{color,colortbl}
\usepackage{xcolor}
\usepackage{multirow}

\newcommand{\gdot}{\dot{\gamma}}
\newcommand{\be}{\begin{equation}}
\newcommand{\ee}{\end{equation}}

\usepackage[colorlinks, citecolor=blue, urlcolor=red, linkcolor=blue]{hyperref}

\newcommand{\smf}[1]{\textcolor{red}{#1}}

\begin{document}

\title{Constitutive model for the rheology of biological tissue}

\author{Suzanne M. Fielding}
\affiliation{Department of Physics, Durham University, Science Laboratories,  South Road, Durham DH1 3LE, UK}
\author{James O. Cochran}
\affiliation{Department of Physics, Durham University, Science Laboratories,
  South Road, Durham DH1 3LE, UK}
\author{Junxiang Huang}
\affiliation{Department of Physics, Northeastern University, MA 02115, USA}
\author{Dapeng Bi}
\affiliation{Department of Physics, Northeastern University, MA 02115, USA}
\author{M. Cristina Marchetti}
\affiliation{Department of Physics, University of California, Santa Barbara, CA, USA}

\begin{abstract}

The rheology of biological tissue is key to processes such as embryo development, wound healing and cancer metastasis. Vertex models of confluent tissue monolayers have uncovered a spontaneous liquid-solid transition tuned by cell shape; and a shear-induced solidification transition of an initially liquid-like tissue.  Alongside this jamming/unjamming behaviour, biological tissue also displays an inherent viscoelasticity, with a slow time and rate dependent mechanics. With this motivation, we combine simulations and continuum theory to examine the rheology of the vertex model in nonlinear shear across a full range of shear rates from quastistatic to fast,  elucidating its nonlinear stress-strain curves after the inception of shear of finite rate, and its steady state flow curves of stress as a function of strain rate. We formulate a rheological constitutive model that couples cell shape to flow and captures both the tissue solid-liquid transition and its rich linear and nonlinear  rheology. 

\end{abstract}

\maketitle

The rheology of biological tissue is crucial to processes such as morphogenesis, wound healing and cancer metastasis.  On short timescales, tissues withstand stress in a solid-like way. On longer timescales, they reshape via internally active processes such as cell shape change, rearrangement, division and death~\cite{guirao2015unified,etournay2015interplay}. Tissues are thus viscoelastic~\cite{forgacs1998viscoelastic}. Power law stress relaxation~\cite{Hoffman2006,khalilgharibi2019stress}  and slow oscillatory cell displacements~\cite{Pruitt2018} after straining underline their rate dependent mechanics. Tissues furthermore undergo spontaneous solid-liquid  transitions~\cite{bi_nphys_2015,Park_NMAT_2015,bi2016motility,
malinverno2017endocytic,mongera2018fluid} driven by both active processes, such as fluctuations of cell-edge tensions, motility and alignment, and geometric constraints~\cite{lawson2021jamming}, with important implications for morphogenesis and cancer progression.
Nonlinear rheological response to tensile stretching includes stiffening~\cite{fernandez2006master} or fluidization~\cite{trepat_fredberg_stretch_nature_2007} of single cells, and stiffening then rupture of tissue monolayers~\cite{Harris_PNAS_stretch}. Internal activity can likewise induce nonlinear phenomena such as  superelasticity~\cite{latorre2018active} and fracture~\cite{prakash2021motility}. 

Understanding tissue rheology theoretically is thus of major importance. 
Well studied vertex and Voronoi models~\cite{Nagai_PMB_2001,Farhadifar_CB_2007,bi2016motility} of confluent tissue, with no gaps between cells, represent a 2D tissue monolayer as a tiling of polygonal cells.  They capture a density-independent solid-liquid transition tuned by a parameter characterising the target cell shape, which in turn embodies the competition between cortex contractility and cell-cell adhesion~\cite{bi_nphys_2015,Park_NMAT_2015,bi2016motility}.
Vertex models have also been used to study the linear mechanics of tissues~\cite{moshe2018geometric,Rastko_2021,hernandez2022anomalous}, and their response to nonlinear stretch~\cite{Merzouki_vm_strain} and shear~\cite{Popovic_2021,duclut2021nonlinear,PicaCiamarra_rheology,huang2022shear}. Recently, vertex model simulations of a tissue that is fluid-like in zero shear demonstrated a shear-induced rigidity transition above a critical strain, applied quasistatically~\cite{huang2022shear}.

While vertex models and other mesoscopic models have played an important role in advancing our understanding of tissue mechanics, it is also helpful to develop coarse grained continuum rheological constitutive models. 
Early work formulated a continuum model that couples cell shape and cell motility, capturing some of the glassy dynamics of tissue~\cite{czajkowski2018hydrodynamics}.  
Inspired by early hydrodynamic theories of active fluids and gels~\cite{marchetti2013hydrodynamics,prost2015active}, continuum constitutive models have been developed to characterize the role of cell shape change, rearrangements, division and death in morphogenesis~\cite{ranft2010fluidization,etournay2015interplay,duclut2021nonlinear,dye2021self,grossman2022instabilities,ishihara2017cells,murisic2015discrete}. 

Still lacking, however, is a continuum hydrodynamic constitutive model capable of describing both  the spontaneous solid-liquid transition of confluent tissues and its rheological response to external deformation and flow.
Inspired by  mean-field theories of cell-shape driven transitions~\cite{czajkowski2018hydrodynamics,hernandez2022anomalous,huang2022shear} and by fluidity models of the rheology of dense soft suspensions~\cite{picard2002simple}, we introduce such a model.

The key new insights of our approach are as follows. First, we distinguish the role of {\em geometric frustration} (encoded in the cell perimeter $p$), from that of T1 {\em topological rearrangements} (encoded in our fluidity variable $a$). The former is key to the zero-shear liquid-solid transition and (when coupled to our orientation tensor $\sigma_{ij}$)  strain stiffening  at small to modest imposed strains~\cite{huang2022shear}. The latter cause the plasticity associated with the stress overshoot at imposed strains $O(1)$, and the ultimate steady flowing state. Second, in modeling the geometric frustration, we distinguish a tensor characterizing individual cell shape  (of which $p$ is the trace), and a tensor characterizing the average cell orientation at the tissue scale~\cite{czajkowski2018hydrodynamics}.

We furthermore submit this new continuum model to stringent comparison with 
simulations 
across a full range of shear rates from quasi-static to fast. 
We demonstrate our continuum model to capture both the zero-shear solid-liquid transition and strain stiffening transitions reported in Ref.~\cite{huang2022shear}, the full nonlinear stress vs. strain behavior after the inception of shear, and the steady state flow curves of stress vs. shear rate.

{\it Vertex model simulations ---}  The vertex model~\cite{Nagai_PMB_2001,Farhadifar_CB_2007} represents the tightly packed confluent cells of a 2D tissue monolayer as $c=1\cdots N_c$ polygons that tile the plane. Each cell is defined by the location of its $n_c=1\cdots \nu_c$ vertices, with any two neighbouring vertices $\alpha$ and $\beta$ connected by an edge of length $\ell_{\alpha\beta}$. 
The elastic energy of the tissue is controlled by the interplay of pressure within each cell and tension along the cell edges. Assuming the cell-edge tension per unit length is uniform across the tissue, the energy can be written as
\begin{equation}
    E=\frac12\sum_c\left[\kappa_A(A_c-A_{c0})^2+\kappa_P(P_c-P_{c0})^2\right]\;,
    \label{eq:E}
\end{equation}
where each cell experiences an energy cost for deviation of its area $A_c$ and perimeter $P_c$ from target values $A_{c0}$ and $P_{c0}$, with area and perimeter stiffness $\kappa_{\rm A}$ and $\kappa_P$. 
The first term on the RHS  models 3D cell volume incompressibility via an effective 2D area elasticity~\cite{Farhadifar_CB_2007,Staple_PJE_2010}. The second 
describes the competition between cell cortical contractility and adhesion between neighbouring cells in controlling cell-edge tension and perimeter~\cite{Staple_PJE_2010,Farhadifar_CB_2007,bi_nphys_2015}.

We denote by $\Vec{F}_n=-\frac{\delta E}{\delta\Vec{x}_n}$ the total force on the $n^{th}$ vertex of the tiling at position $\Vec{x}_n$ due to interactions with all other vertices. In an applied shear of rate $\gdot$, with flow direction $x$ and shear gradient $y$, we assume over-damped dynamics with drag  $\zeta$,
%
$\frac{d\Vec{x}_n}{dt}=\zeta^{-1}\Vec{F}_n+\gdot y_n \Vec{\hat{x}}$,
%
with Lees-Edwards periodic boundary conditions. The cells also undergo T1  topological neighbor exchanges that allow the tissue to plastically relax stresses~ \cite{bi2016motility,yang2017correlating,Das_T1,ourSI}.

To focus on amorphous tissue structures, we simulate a $50:50$ bidisperse  tiling of 
$N_c=4096$  
cells of   target areas $A_0=1, 1.4$, which sets our length unit. 
We adjust $P_{c0}$ for the two cell populations to maintain the target cell shape $p_0=P_{c0}/\sqrt{A_{c0}}$ the same for all cells. We choose units in which  $\kappa_A=1$ and $\zeta=1$ and set $\kappa_p=1.0$ throughout. We vary $p_0$ and the imposed  shear rate $\gdot$. 
As an initial condition, we seed a planar Voronoi tiling then evolve the above dynamics to steady state in zero shear.  
At time $t=0$, we switch on shear and measure the shear stress  $\Sigma_{ij}(t)=\tfrac{1}{N}\sum_{n=1}^N F_{ni}x_{nj}$, where the sum is over all $N$ vertices in the tiling, and the mean  cell perimeter $p(t)=\tfrac{1}{N_c}\sum_{c=1}^{N_c}p_c$. Denoting by $\Vec{t}^{n_c}$ the unit vector along the  edge of length $l_{n_c}$ between the $n_c th$ and $(n_c +1)th$ vertices of cell $c$,  we define  a single-cell shape tensor $\sigma^c_{ij}=\tfrac{1}{\nu_c}\sum_{n=1}^{\nu_c}l_{n_c}t^{n_c}_it^{n_c}_j$, where the sum is over the $\nu_c$ vertices of the $c$-th cell,  and the tissue-scale averaged orientation tensor $\sigma_{ij}=\tfrac{1}{N_c}\sum_{c=1}^{N_c}\sigma^c_{ij}$. We use the same notation $\Sigma_{ij},\sigma_{ij},p$ for the counterpart coarse-grained quantities in our constitutive model below.

In the absence of external stress, the vertex model  exhibits a liquid solid transition as a function of the target shape $p_0$~\cite{bi_nphys_2015,yan_bi_rigidity}. For $p_0<p_0^*$ the energy barriers to T1 transitions are finite and the system is a solid with a finite zero-frequency linear shear modulus. At the critical value $p_0^*$, the mean energy barrier for T1 transitions vanishes, giving liquid response for $p_0>p_0^*$. For our bidisperse tiling, $p_0^*=3.85$. For monodisperse disordered polygons  $p_0^*\simeq 3.81$, a value close to that of a regular pentagon~\cite{bi_nphys_2015}. This value is renormalized by motility~\cite{bi2016motility} and by cell alignment with local spontaneous shear~\cite{wang2020anisotropy}. 
It was recently realized that this transition has a geometric origin associated with the underconstrained nature of the energy in Eq.~\ref{eq:E}~\cite{moshe2018geometric,merkel2018geometrically,hernandez2022anomalous}.
For regular hexagons the transition occurs at the isoperimetric value $p_{\rm iso}=\sqrt{8\sqrt{3}}\simeq3.722$. Below this value  it is not possible to satisfy both target area and perimeter and the ground state has $p=p_0^*$ and finite energy. This is the solid or incompatible state. For $p_0>p_{\rm iso}$ there is a family of zero energy area and perimeter preserving ground states, with $p=p_0$. The system can accommodate an externally applied linear shear by adjusting its shape within this degenerate manifold~\cite{hernandez2022anomalous}. The compatible system is therefore a liquid with zero shear modulus, although it stiffens and acquires rigidity at finite strains~\cite{huang2022shear}.

{\it Constitutive model ---}  
We now construct a continuum model that accounts for the mean-field liquid-solid transition, and also captures the key rheological features of the vertex model:
(i) reversibility of linear response to small strains,
(ii) strain stiffening at intermediate strains, (iii) plastic relaxation at larger strains, due to T1 cell rearrangements, and (iv) a yield stress in the steady state flow curve $\Sigma(\gdot)$, as obtained in Ref.~\cite{huang2022shear}. Although our model below is cast in frame invariant form, capable of addressing any flow, we focus on response to simple shear, to compare with our vertex model simulations.

We assume dynamics of the cell perimeter governed by:
\begin{eqnarray}
\label{eqn:perimeter}
    \dot{p}+v_k\nabla_kp = \gdot-\frac{1}{\tau_{\rm p}}(p-p_0)(p-p_0^*-\alpha\sigma_{ij}\sigma_{ij})\;,
    \end{eqnarray}
with $\alpha$ and $\tau_{\rm p}$ constants and invariant strain rate $\gdot=\sqrt{2D_{ij}D_{ij}}$.
In the absence of shear, $p$ relaxes on a timescale $\tau_{\rm p}$ to a steady state that displays a transcritical bifurcation as a function of the target cell perimeter $p_0$, with $p=p_0^*$ in the solid phase $p_0<p_0^*$  and $p=p_0$ in the liquid phase  $p_0>p_0^*$, capturing the liquid-solid transition~\cite{bi_nphys_2015}. The same transcritical structure emerges by writing exact equations for the relaxation of a single 
cell modeled as a regular $n-$sided polygon according to the vertex model dynamics prescribed above. 

In shear, the perimeter is advected by flow and stretched by the shear rate $\gdot$.
In addition, the coupling $\alpha\sigma_{ij}\sigma_{ij}$ 
captures a key intuition of our approach: that a shear-induced global cell orientation $\sigma_{ij}$ provides an effective mean field that distorts the individual cell's shape $p$ away from its zero-shear value. As a result, in the solid phase $p$ increases relative to its zero shear value $p=p_0^*$ from the outset of straining.
In the liquid phase, $p$ increases relative to its zero shear value $p=p_0$ only after a critical strain amplitude $\gamma_c$, capturing the strain-induced stiffening transition~\cite{huang2022shear}.
The behavior introduced by the coupling of single-cell shape, as quantified by the mean perimeter $p$, to the tissue-scale cell shape $\sigma_{ij}$ is analogous to the influence of cell alignment due to internally generated stresses in  \textit{Drosophila} germband extension~\cite{wang2020anisotropy}. Indeed, the form of coupling of $p$ to $\sigma_{ij}$ in Eqn.~\ref{eqn:perimeter} is justified both by experiment~\cite{wang2020anisotropy} and mean field theory~\cite{huang2022shear,hernandez2022anomalous}.

The cell orientation tensor is taken to obey an evolution equation of the widely used Maxwellian form,
\begin{equation}
\label{eqn:Maxwell}
 \dot{\sigma}_{ij}+v_k\nabla_k\sigma_{ij}=\sigma_{ik}K_{kj}+K_{ki}\sigma_{kj}+2D_{ij}-a\sigma_{ij}\;,
\end{equation}
where $K_{ij}=\partial_j v_i$ is the strain rate tensor and  $D_{ij}=\tfrac{1}{2}(K_{ij}+K_{ji})$.  The last  term in Eq.~\ref{eqn:Maxwell} describes  plastic relaxation. It vanishes in linear response (small strains), where $a=0$ (see below), allowing  the orientation tensor $\sigma_{ij}$ to build linearly and reversibly with strain, as expected in the absence of plastic T1 events. 

Consistent with previous studies of the vertex model~\cite{Farhadifar_CB_2007,yang2017correlating,hernandez2022anomalous} we write the deviatoric stress tensor
\begin{equation}
\label{eqn:stress}
 \tilde\Sigma_{ij}=C(p-p_0)\left(\sigma_{ij}-\frac12\delta_{ij}\sigma_{kk}\right)\;.
\end{equation}
Here $C$ is constant and $p_0$ the target cell perimeter. In linear response (small strains), the effective modulus $G_0=C(p-p_0)$ is non-zero in the solid phase, where $p>p_0$, and zero in the liquid phase, where $p=p_0$.

Were the factor $a$  on its RHS a constant inverse relaxation time, Eq.~\ref{eqn:Maxwell} would be the widely used Maxwell model, capturing viscoelasticity, but not the irreversible plasticity of T1 events. To model plasticity, we take $a$ to be a fluidity-like variable~\cite{picard2002simple} with dynamics:
\begin{equation}
\label{eqn:fluidity}
    \dot{a}+v_k\nabla_{k}a=\gdot\left[-a+f(\gdot)\right],
\end{equation}
with  $f(\gdot)=\beta\gdot/(1+\tfrac{1}{2}\tau_0\gdot)$,
in which $\beta$ is constant and $\tau_0$ a microscopic time. As suited to an athermal tissue, with no relaxation events induced by temperature or activity (no cell motility, division or death), this is a purely strain-driven dynamics. In linear response, $a=0$, giving a reversible dependence of $\sigma_{ij}$ on strain. In weak shear, $a$ builds on a strain $O(1)$ to  model the plasticity of T1 events via the final term in Eq.~\ref{eqn:Maxwell}. In steady weak shear $a=f(\gdot)\approx \beta\gdot$, giving a divergent relaxation time $1/a$ as $\gdot\to 0$, and a yield stress in the steady state flow curve. 

 We explore  different values of shear rate, $\gdot$, and the target perimeter $p_0$ relative to the transition  $p_0^*$. (See Appendix for model parameters.)
\begin{figure}[t]
    \includegraphics[width=9.0cm]{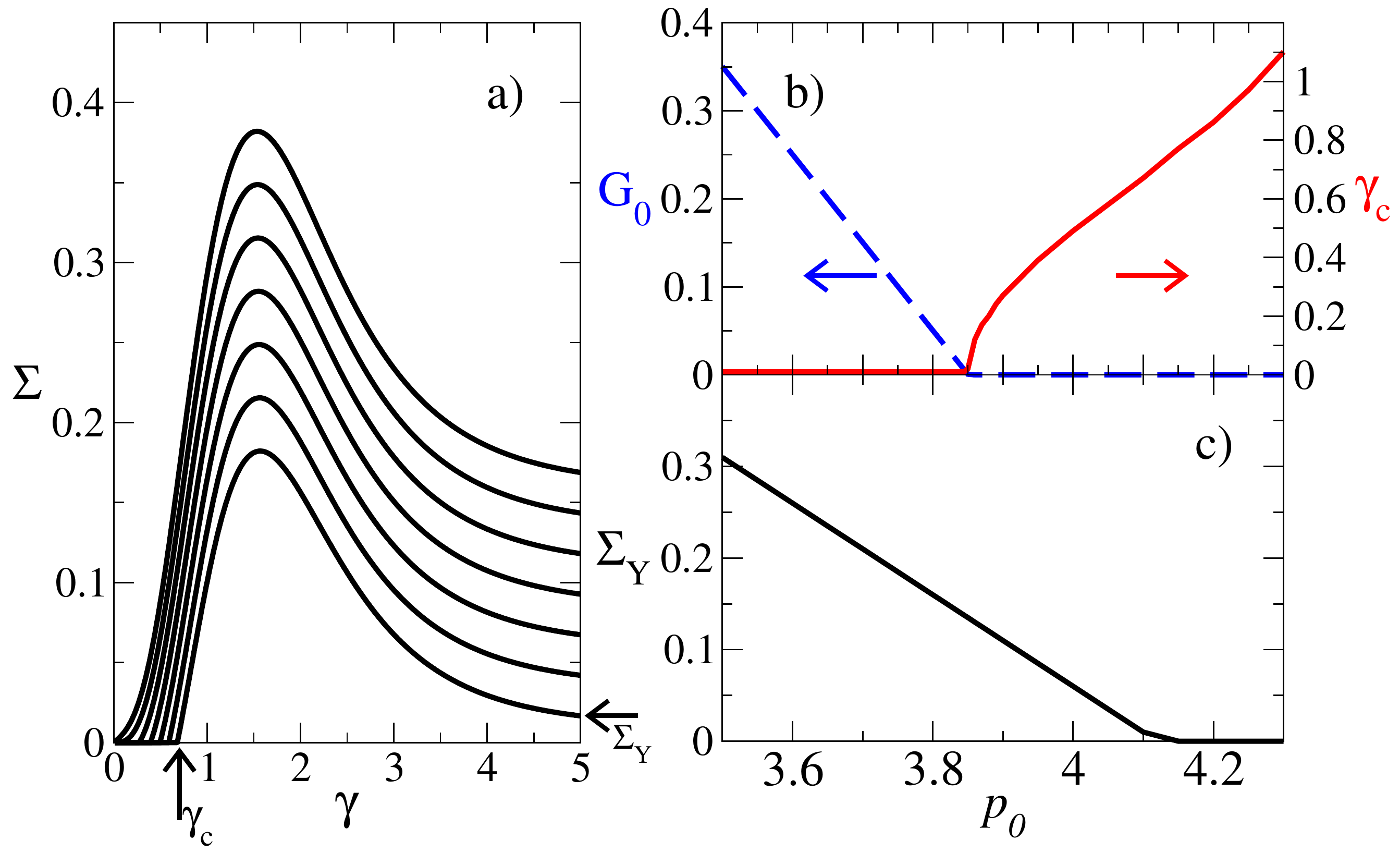}
    \caption{Constitutive model in slow shear, $\gdot=10^{-6}$, probing the quasistatic limit $\gdot\to 0$. {\bf (a)} Stress vs. strain after the switch-on of shear, target perimeter $p_0=3.80,3.85\cdots 4.10$ in curves downwards. {\bf (b)} Linear elastic modulus $G_0=d\Sigma/d\gamma|_{\gamma=0}$ (dashed line) and strain $\gamma_c$ at the shear-induced solidification transition (solid line).  {\bf (c)} Shear stress in the limit of steady shear $\gamma\to\infty$. }
    \label{fig:quasistatic}
\end{figure}
We prescribe as initial condition to shear a perimeter  $p(t=0)$ equal to its steady state value in zero shear,  an orientation tensor $\sigma_{ij}(t=0)=0$, and fluidity $a(t=0)=0$.
We then switch on  a simple shear $K_{ij}=\gdot \delta_{iy}\delta_{jx}$ at time $t=0$  and track the evolution of  $p, \sigma_{xy}$ and $\Sigma_{xy}$ as a function of  time $t$ or equivalently (to within a constant factor $\gdot$)  accumulating strain $\gamma=\gdot t$. 
Hereafter we drop the $xy$ subscript, writing  $\sigma_{xy}=\sigma$ and $\Sigma_{xy}=\Sigma$.

{\it Results ---}  Our constitutive model captures the liquid-solid transition as a function of target cell shape in zero shear~\cite{bi_nphys_2015} and the shear-induced rigidity transition of the liquid-like tissue, above a critical  shear strain, applied quasistatically $\gdot\to 0$~\cite{huang2022shear}. See Fig.~\ref{fig:quasistatic}a, which shows
the shear stress $\Sigma$ vs. strain $\gamma$ in shear at rate $\gdot=10^{-6}$.
At small strains, just after the inception of shear, the modulus $G_0=d\Sigma/d\gamma|_{\gamma=0}$ is finite (solid-like) for $p_0<p_0^*$ but zero (liquid-like) for $p_0>p_0^*$ (Fig.~\ref{fig:quasistatic}b, dashed line).
In the liquid phase, the stress $\Sigma$ and slope $d\Sigma/d\gamma$ first become non-zero above a nonlinear critical shear strain $\gamma_c$, heralding a strain-induced stiffening transition (solid line in Fig.~\ref{fig:quasistatic}b, defined as the strain at which the stress first exceeds $10^{-5}$ at any  $p_0$.)

\begin{figure}[t]
    \includegraphics[width=9.0cm]{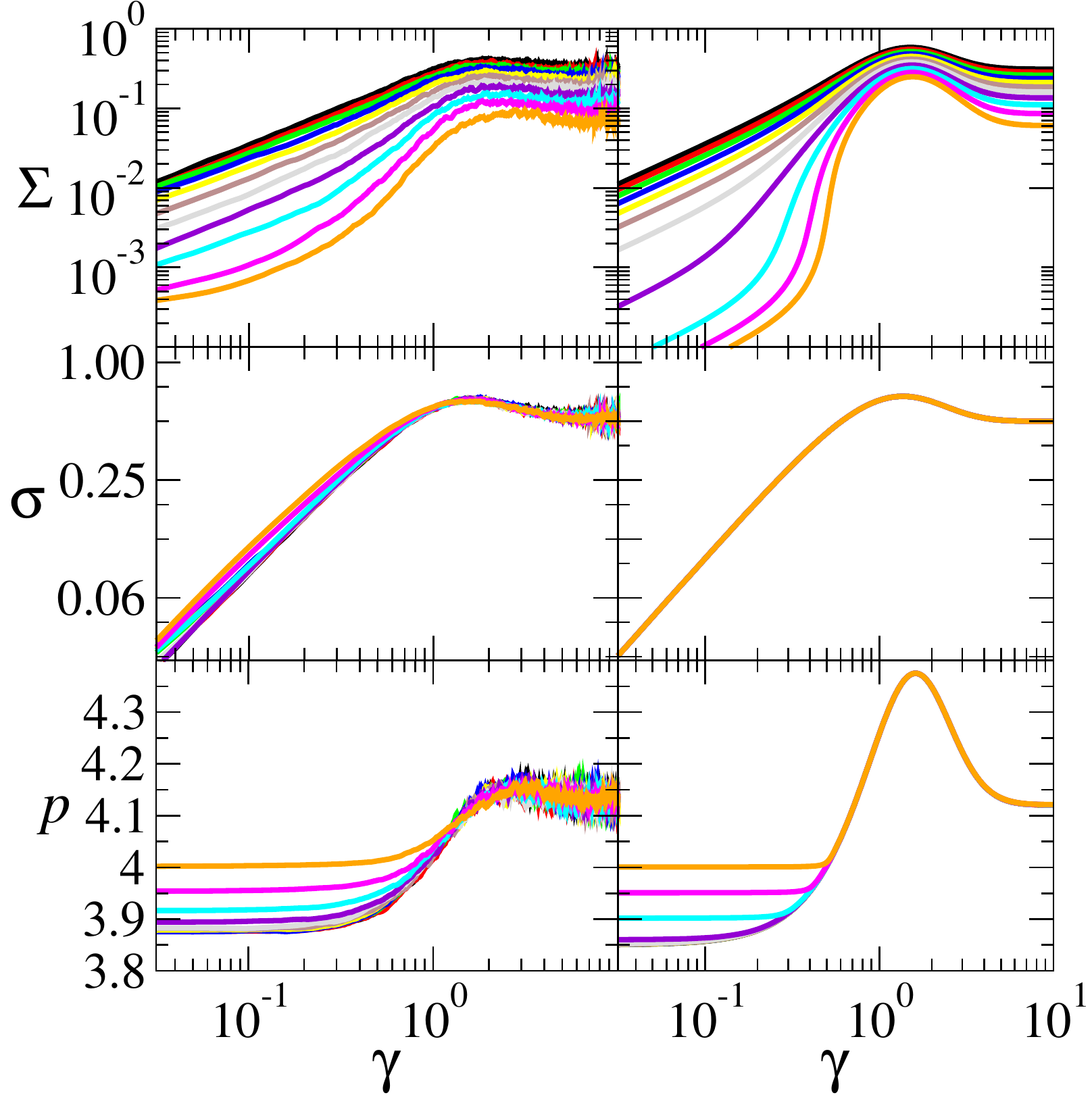}
    \caption{Rheological behaviour of the vertex model {\bf (left)} and constitutive model {\bf (right)} in shear startup at a shear rate $\gdot=10^{-3}$ for values of the target perimeter $p_0=3.50,3.55,3.60\cdots 4.00$ (in black, red, green $\cdots$ orange curves downwards; curve for $p_0^*=3.85$ in purple). Shown is the evolution of the shear stress {\bf (top)}, shear component of the orientation tensor {\bf (middle)} and cell perimeter {\bf (bottom)} as a function of accumulating strain $\gamma=\gdot t$. } 
    \label{fig:startup}
\end{figure}

Having explored quasistatic shear, we now consider nonlinear shear flow across a full range of shear rates from quasi-static to fast.
The evolution of $\Sigma,\sigma$ and $p$ as a function of strain since the inception of shear is shown in Fig.~\ref{fig:startup},
for  a range of $p_0$ below and above $p_0^*$. The left column shows the results of vertex model simulations. The right shows the predictions of our constitutive model, which performs well in capturing all the qualitative features of the  simulations.

At small strains, just after shearing starts, the effective modulus $G_0=d\Sigma/d\gamma|_{\gamma=0}$ is finite in the solid phase, $p_0<p_0^*$, but small in the liquid phase, $p_0>p_0^*$. Indeed,  repeating the simulations for progressively lower strain rates $\gdot\to 0$ in the solid phase, $G_0$ tends to a non-zero constant, $G_0(p_0,\gdot\to 0)$, consistent with the quasistatic results discussed above. In the liquid phase,  $G_0\to 0$  as $\gdot\to 0$,
again consistent with the quasistatic results. 

At higher strains, $\gamma=O(1)$,  strain stiffening is observed: the slope of $\Sigma$ vs $\gamma$ increases with increasing $\gamma$. This is particularly pronounced in the liquid phase, $p_0>p_0^*$, where the effective modulus $d\Sigma/d\gamma$ was very small at small strains (tending to zero as $\gdot\to 0$, as just discussed), but becomes appreciable after a strain $\gamma=O(1)$ (even in the limit $\gdot\to 0$). After this regime of strain stiffening, the stress overshoots slightly before declining to a constant in the final state of steady flow.

\begin{figure}[t]
    \includegraphics[width=9.0cm]{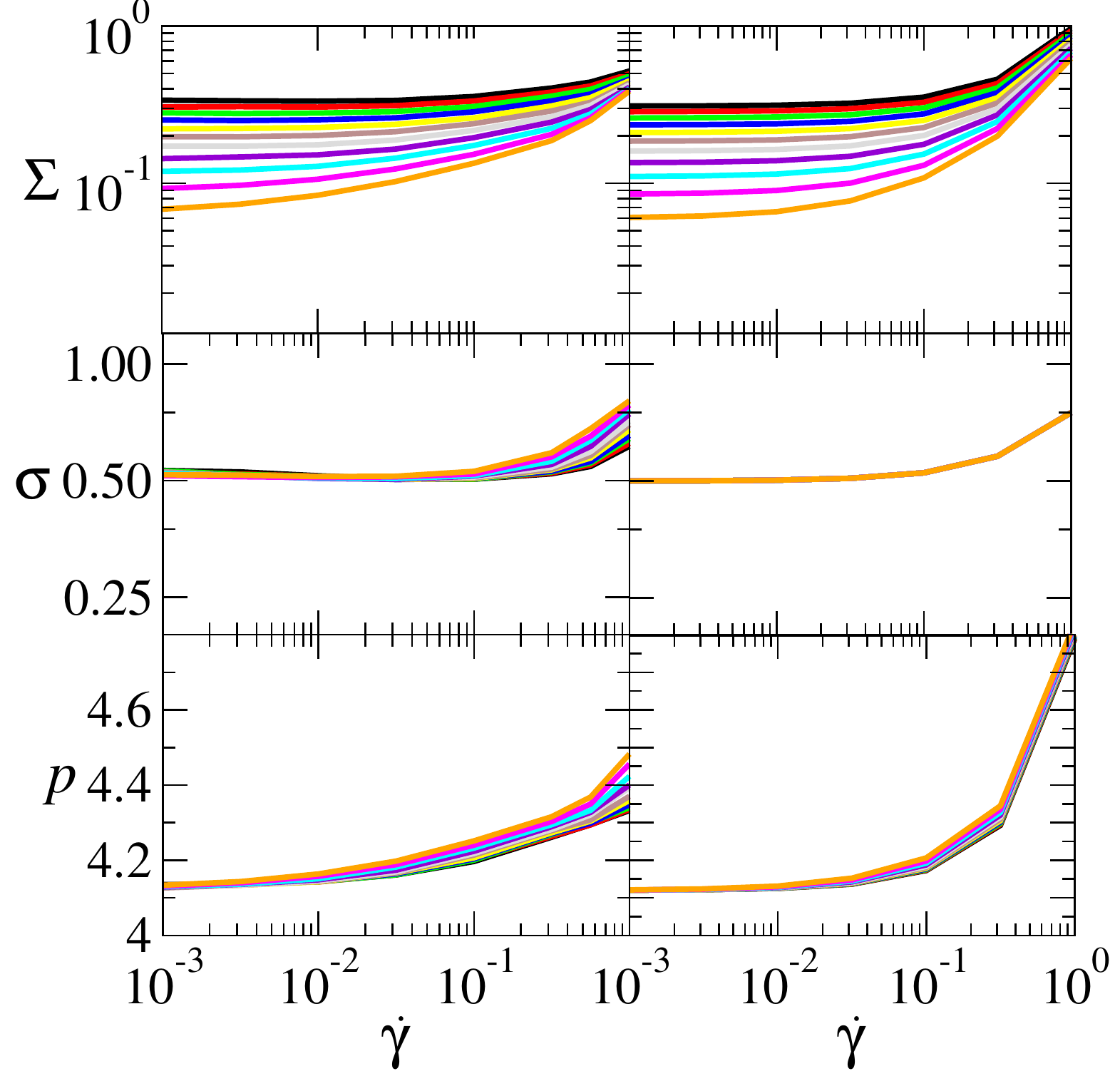}
    \caption{Steady state ($t\to\infty$) dependence of the shear stress {\bf (top)}, shear component of the orientation tensor {\bf (middle)} and cell perimeter {\bf (bottom)} for the same values of the target perimeter $p_0$ as in Fig.~\ref{fig:startup}, with the same line colour coding. Results are shown for the vertex model in the {\bf left} column and the constitutive model in the {\bf right} column.}
    \label{fig:flowCurves}
\end{figure}

This rich behaviour  is readily understood within our simple constitutive model.
The initial fluid-like behaviour 
for $p_0>p_0^*$ arises because  $p=p_0$  before shearing commences, giving zero effective modulus in Eq.~\ref{eqn:stress}. As strain increases, tissue deformation is captured by the growth of $\sigma$, which in turn yields  an increase of $p$ relative to its equilibrium value due to the coupling term in $\alpha$ in Eq.~\ref{eqn:perimeter}. This is also responsible for the less pronounced strain stiffening  in the solid phase, $p_0<p_0^*$. The subsequent overshoot in stress $\Sigma$ (and  perimeter $p$) at larger strains is caused by the overshoot in the cell orientation $\sigma$ seen in the middle panels of Fig.~\ref{fig:startup}. The stress decline after overshoot arises in the vertex model from plastic relaxation via T1 events, an effect captured in the constitutive model via an increase of fluidity $a$ with shear. The tissue shape tensor $\sigma$ is essentially independent of $p_0$ in the vertex model (at low strain rates), consistent with the lack of any coupling of the evolution equation for $\sigma_{ij}$ to $p$ in the constitutive model.

 At long times, $t\to\infty$, after many strain units $\gamma=\gdot t\to\infty $, a state of final plastic flow is reached in which each of  $\Sigma,\sigma$ and $p$ attains a steady value. This is reported as a function of $\gdot$ in Fig.~\ref{fig:flowCurves},  for the vertex model (left column), and constitutive model (right), with good semi-quantitative agreement. 
In rheological parlance, the steady state relationship $\Sigma=\Sigma(\gdot)$ is termed the ``flow curve''. 
The vertex model flow curves show a dynamical yield stress: a non-zero limiting intercept $\lim_{\gdot\to 0}\Sigma(\gdot)=\Sigma_{\rm Y}\neq 0$. 
Importantly, this is true both for $p_0<p_0^*$ and for $p_0>p_0^*$: whereas liquid and solid states are distinct and separated by a transition at small strains, 
in steady nonlinear shear, however slow, the vertex model displays a non-zero yield stress up to a larger $p_0=p_0^{**}>p_0^*$~\cite{huang2022shear}, as also seen in Fig.~\ref{fig:quasistatic}. 
This is easily understood within our constitutive model.
In steady shear, Eq.~\ref{eqn:fluidity} predicts the fluidity $a=f(\gdot)=\beta\gdot/(1+\tfrac{1}{2}\tau_0\gdot)$. Combining with
 Eq.~\ref{eqn:Maxwell} for the orientation gives $\sigma=\gdot/a=\tfrac{1}{\beta}(1+\tfrac{1}{2}\tau_0\gdot)$.
Were we to assume $p-p_0=1$, independent of strain rate, we would obtain a flow curve  $\Sigma(\gdot)=\tfrac{C}{\beta}(1+\tfrac{1}{2}\tau_0\gdot)$, with a yield stress $\sigma_Y=\tfrac{C}{\beta}$ as $\gdot\to 0$ and Newtonian behaviour $\Sigma\propto\gdot$ as $\gdot\to\infty$. The actual flow curve is modified somewhat in comparison, due to the strain rate dependence of $p-p_0$. Importantly, however, it retains a yield stress because $p\neq p_0$ in steady flow, even in the limit $\gdot\to 0$: the perimeter is always strongly perturbed from its unsheared value, due to the coupling $\alpha\sigma_{ij}\sigma_{ij}$ in Eq.~\ref{eqn:perimeter}. 
Intuitively, the key effect of a steady shear, even when applied quasistatically, is to deform cells away from their target shape such that they carry a stress and the liquid phase seen at small strains is destroyed.

{\em Conclusions ---} 
We have presented a continuum constitutive model for the rheology of confluent 2D biological tissue and demonstrated it to capture the rich rheophysics seen in simulations of the vertex model under applied shear. This  includes strain-stiffening of the liquid above a critical strain, a stress overshoot at larger strains due to the plasticity of T1 rearrangements, and a finite yield stress in steady shear, even in the (zero-shear) liquid phase.   Our model includes the effects of cell shape change and rearrangements on mechanical behaviour, and will provide a useful phenomenological framework for modeling the rheology of biological tissue. Elucidating its  predictions in deformation protocols besides simple shear is left to future work, as are extensions to incorporate other  active processes such as cell motility, division and death.
\vspace{0.1in}

{\it Acknowledgements ---} 
M.C.M. thanks Mike Cates for an early discussion that motivated this project and Arthur Hernandez for many illuminating discussions. S.M.F. received funding from the European Research Council (ERC) under the European Union's Horizon 2020 research and innovation programme (grant agreement No. 885146). M.C.M. was supported by the US National Science Foundation Grant No. DMR-2041459. J.H. and D.B.  would like to acknowledge support from the National Science Foundation Grant No. DMR-2046683, the Alfred P. Sloan Foundation and The Human Frontier Science Program. J.O.C was supported by the EPSRC-funded Centre for Doctoral Training in Soft Matter and Functional Interfaces (SOFI CDT - EP/L015536/1).

\section{Appendix: Model and Simulation Parameters}

Model parameters are the modulus-like quantity $C$, the microscopic time $\tau_0$ and the parameter $\beta$ in the function $f$ for the fluidity, the transition value of the target perimeter $p_0^*$, the coupling of perimeter to orientation $\alpha$, and the perimeter relaxation time $\tau_{\rm p}$. We choose units $C=1$ and  $\tau_0=1$, and treat $p_0^*, \alpha, \beta$ and $\tau_{\rm p}$ as fitting parameters in comparing our constitutive model with the vertex model simulations. We have found  $p_0^*=3.85$, $\alpha=0.36$, $\beta=2.0$ and $\tau_{\rm p}=0.1$ to give the best fit. Among these, $p_0^*$ is the value of $p_0$ at the (zero-shear) liquid-solid transition. Accordingly, we set the value of $p_0^*$ in our continuum model to that value  found in our vertex model simulations. $\beta$ sets the quasistatic limit of the shear component of cell orientation tensor, $\lim_{\gdot\to 0}\sigma_{xy}=1/\beta$,
with $\beta=2.0$
in our vertex model simulations. $\alpha$ sets the effective modulus $G(p-p_0)$ in the shear induced solid phase, with $p-p_0 = p_0^*-p_0 + \alpha\sigma_{ij}\sigma_{ij}$ as $\gdot\to 0$, and accordingly sets the  flow curve's yield stress,  $\lim_{\gdot\to 0}\Sigma(\gdot)$. We choose $\alpha$ to give the best fit of the continuum model's yield stress to that of the vertex model simulations. Finally, $\tau_{\rm p}$ 
controls the steepness of the flow curve at high strain rates (where
the vertex model is 
likely to become less reliable 
) and the small finite value $\sim\gdot\tau_{\rm p}$ of the stress before the true quasistatic strain-stiffening transition.

The numerical timestep is $Dt=\tilde{Dt} l_{\rm min}/F_{\rm max}$ with $l_{\rm min}$ the minimum edge length, $F_{\rm max}$ the maximum vertex force and $\tilde{Dt}=0.01$. T1 events are triggered below a critical edge length $l_{\rm c}=0.01$. 

\section*{Supplemental Material}

\subsection*{Vertex model simulations}

The vertex model~\cite{Nagai_PMB_2001,Farhadifar_CB_2007} represents the confluent cells of a tissue monolayer via $c=1\cdots N_c$ polygons that tile the plane. Each cell is defined by the location of its $n_c=1\cdots \nu_c$ vertices, with any two neighbouring vertices connected by an edge. Each vertex belongs to three neighbouring cells (or transiently four, during a T1 event, see below), with three (or four) edges stemming from it accordingly. Each edge belongs to two neighbouring cells. 

\begin{figure}[!b]
    \includegraphics[width=9.0cm]{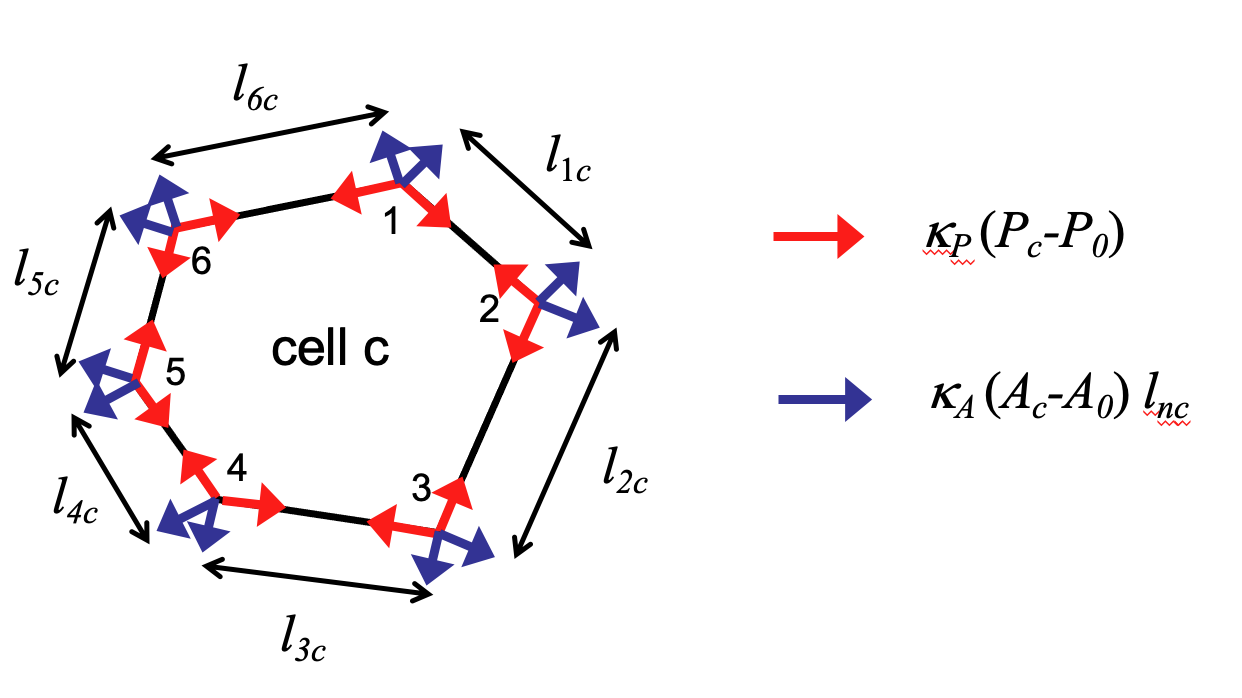}
    \caption{Sketch of vertex model forces.}
    \label{fig:sketch}
\end{figure}

Consider the $n_c th$ and $(n_c+1)th$ vertices of cell $c$. Cell $c$ contributes to  these two vertices an equal and opposite force of magnitude, $\kappa_{\rm P}(P_c-P_0)$, acting as a tension along the edge that connects them. This models a competition between cell cortical contractility and adhesion between neighbouring cells~\cite{Staple_PJE_2010}, with $\kappa_P$ an elastic constant \cite{Farhadifar_CB_2007}, $P_c$ the cell perimeter and $P_0$ its target value~\cite{bi_nphys_2015}. Cell $c$ furthermore contributes to the same two vertices a force of magnitude $\kappa_{\rm A}(A_c-A_0)l_{nc}$,  acting as a pressure in the direction of the edge normal outwards from cell $c$, with $l_{nc}$ the length of the edge connecting the vertices. Physically, this models 3D cell volume incompressibility via an effective 2D area elasticity of constant $\kappa_A$, with $A_c$ the cell area and $A_0$ its target value~\cite{Farhadifar_CB_2007,Staple_PJE_2010}. Fig.~\ref{fig:sketch} shows a sketch of these forces.  Each of the two vertices also belongs to two  other neighbouring cells (or three, during T1 events), which contribute forces likewise.

For the $nth$ vertex of all $N$ in the tiling, we denote the sum of the forces from each of its associated cell edges by $F_n$. In shear of rate $\gdot$, with flow direction $x$ and shear gradient $y$, the vertex position $\Vec{x}_n$ obeys over-damped dynamics with drag coefficient $\zeta$ as a function of time $t$:
\be
\frac{d\Vec{x}_n}{dt}=\frac{1}{\zeta}\Vec{F}_n+\gdot y_n \Vec{\hat{x}},
\ee
with Lees-Edwards periodic boundary conditions.

Consider the vertex at the junction between cells $\alpha\beta\gamma$ and a neighbouring vertex between cells $\alpha\beta\delta$. When the edge connecting these vertices shrinks below a small length $l_{\rm c}$, a $T1$ transition removes these two vertices and replaces them with new ones at the junctions of cells $\beta\gamma\delta$ and $\alpha\gamma\delta$, conferring plastic cell rearrangement.

\subsection*{Componentwise constitutive equations in simple shear}

In the main text, we presented our constitutive model in tensorial, frame-invariant form, capable of addressing any imposed deformation or flow protocol. Here we extract the components of those equations relevant to homogeneous imposed simple shear flow with  $K_{ij}=\gdot \delta_{iy}\delta_{jx}$.

The cell perimeter evolves according to:
\begin{eqnarray}
\label{eqn:perimeterC}
    \frac{dp}{dt} &=& \gdot+\frac{1}{\tau_{\rm p}}(p_0-p_0^*-\alpha\sigma_{ij}\sigma_{ij})(p-p_0^*-\alpha\sigma_{ij}\sigma_{ij})\nonumber\\
    & &-\frac{1}{\tau_{\rm p}}(p-p_0^*-\alpha\sigma_{ij}\sigma_{ij})^2\;,
    \end{eqnarray}
where $\alpha$ and $\tau_{\rm p}$ are constants, and the trace of the cell orientation tensor $\sigma_{ii}=\sigma_{xx}+\sigma_{yy}$.

The orientation tensor obeys an evolution equation of the widely used Maxwellian form, which componentwise is written as:
\begin{eqnarray}
\label{eqn:MaxwellC}
 \frac{d\sigma_{xy}}{dt}&=&\gdot(1+\sigma_{yy})-a\sigma_{xy}\nonumber\\
 \frac{d\sigma_{xx}}{dt}&=&2\gdot\sigma_{xy}-a\sigma_{xx}\nonumber\\
 \frac{d\sigma_{yy}}{dt}&=&-a\sigma_{yy}.
\end{eqnarray}
The shear stress
\begin{equation}
\label{eqn:stressC}
 \Sigma_{xy}=C(p-p_0)\sigma_{xy},
\end{equation}
where $C$ is constant and $p_0$ the target cell perimeter. 

To model plasticity, we take the quantity $a$ in the evolution of the orientation tensor to be a fluidity-like variable~\cite{picard2002simple} with its own dynamics:
\begin{equation}
\label{eqn:fluidityC}
    \frac{da}{dt}=\gdot\left[-a+f(\gdot)\right],
\end{equation}
with and $f(\gdot)=\beta\gdot/(1+\tfrac{1}{2}\tau_0\gdot)$,
in which $\beta$ is constant and $\tau_0$ a microscopic time.


\begin{thebibliography}{43}%
\makeatletter
\providecommand \@ifxundefined [1]{%
 \@ifx{#1\undefined}
}%
\providecommand \@ifnum [1]{%
 \ifnum #1\expandafter \@firstoftwo
 \else \expandafter \@secondoftwo
 \fi
}%
\providecommand \@ifx [1]{%
 \ifx #1\expandafter \@firstoftwo
 \else \expandafter \@secondoftwo
 \fi
}%
\providecommand \natexlab [1]{#1}%
\providecommand \enquote  [1]{``#1''}%
\providecommand \bibnamefont  [1]{#1}%
\providecommand \bibfnamefont [1]{#1}%
\providecommand \citenamefont [1]{#1}%
\providecommand \href@noop [0]{\@secondoftwo}%
\providecommand \href [0]{\begingroup \@sanitize@url \@href}%
\providecommand \@href[1]{\@@startlink{#1}\@@href}%
\providecommand \@@href[1]{\endgroup#1\@@endlink}%
\providecommand \@sanitize@url [0]{\catcode `\\12\catcode `\$12\catcode
  `\&12\catcode `\#12\catcode `\^12\catcode `\_12\catcode `\%12\relax}%
\providecommand \@@startlink[1]{}%
\providecommand \@@endlink[0]{}%
\providecommand \url  [0]{\begingroup\@sanitize@url \@url }%
\providecommand \@url [1]{\endgroup\@href {#1}{\urlprefix }}%
\providecommand \urlprefix  [0]{URL }%
\providecommand \Eprint [0]{\href }%
\providecommand \doibase [0]{https://doi.org/}%
\providecommand \selectlanguage [0]{\@gobble}%
\providecommand \bibinfo  [0]{\@secondoftwo}%
\providecommand \bibfield  [0]{\@secondoftwo}%
\providecommand \translation [1]{[#1]}%
\providecommand \BibitemOpen [0]{}%
\providecommand \bibitemStop [0]{}%
\providecommand \bibitemNoStop [0]{.\EOS\space}%
\providecommand \EOS [0]{\spacefactor3000\relax}%
\providecommand \BibitemShut  [1]{\csname bibitem#1\endcsname}%
\let\auto@bib@innerbib\@empty
\bibitem [{\citenamefont {Guirao}\ \emph {et~al.}(2015)\citenamefont {Guirao},
  \citenamefont {Rigaud}, \citenamefont {Bosveld}, \citenamefont {Bailles},
  \citenamefont {{L{\'o}pez-Gay}}, \citenamefont {Ishihara}, \citenamefont
  {Sugimura}, \citenamefont {Graner},\ and\ \citenamefont
  {Bella{\"i}che}}]{guirao2015unified}%
  \BibitemOpen
  \bibfield  {author} {\bibinfo {author} {\bibfnamefont {B.}~\bibnamefont
  {Guirao}}, \bibinfo {author} {\bibfnamefont {S.~U.}\ \bibnamefont {Rigaud}},
  \bibinfo {author} {\bibfnamefont {F.}~\bibnamefont {Bosveld}}, \bibinfo
  {author} {\bibfnamefont {A.}~\bibnamefont {Bailles}}, \bibinfo {author}
  {\bibfnamefont {J.}~\bibnamefont {{L{\'o}pez-Gay}}}, \bibinfo {author}
  {\bibfnamefont {S.}~\bibnamefont {Ishihara}}, \bibinfo {author}
  {\bibfnamefont {K.}~\bibnamefont {Sugimura}}, \bibinfo {author}
  {\bibfnamefont {F.}~\bibnamefont {Graner}},\ and\ \bibinfo {author}
  {\bibfnamefont {Y.}~\bibnamefont {Bella{\"i}che}},\ }\bibfield  {title}
  {\bibinfo {title} {Unified quantitative characterization of epithelial tissue
  development},\ }\href {https://doi.org/10.7554/eLife.08519} {\bibfield
  {journal} {\bibinfo  {journal} {eLife}\ }\textbf {\bibinfo {volume} {4}},\
  \bibinfo {pages} {e08519} (\bibinfo {year} {2015})}\BibitemShut {NoStop}%
\bibitem [{\citenamefont {Etournay}\ \emph {et~al.}(2015)\citenamefont
  {Etournay}, \citenamefont {Popovi{\'c}}, \citenamefont {Merkel},
  \citenamefont {Nandi}, \citenamefont {Blasse}, \citenamefont {Aigouy},
  \citenamefont {Brandl}, \citenamefont {Myers}, \citenamefont {Salbreux},
  \citenamefont {J{\"u}licher},\ and\ \citenamefont
  {Eaton}}]{etournay2015interplay}%
  \BibitemOpen
  \bibfield  {author} {\bibinfo {author} {\bibfnamefont {R.}~\bibnamefont
  {Etournay}}, \bibinfo {author} {\bibfnamefont {M.}~\bibnamefont
  {Popovi{\'c}}}, \bibinfo {author} {\bibfnamefont {M.}~\bibnamefont {Merkel}},
  \bibinfo {author} {\bibfnamefont {A.}~\bibnamefont {Nandi}}, \bibinfo
  {author} {\bibfnamefont {C.}~\bibnamefont {Blasse}}, \bibinfo {author}
  {\bibfnamefont {B.}~\bibnamefont {Aigouy}}, \bibinfo {author} {\bibfnamefont
  {H.}~\bibnamefont {Brandl}}, \bibinfo {author} {\bibfnamefont
  {G.}~\bibnamefont {Myers}}, \bibinfo {author} {\bibfnamefont
  {G.}~\bibnamefont {Salbreux}}, \bibinfo {author} {\bibfnamefont
  {F.}~\bibnamefont {J{\"u}licher}},\ and\ \bibinfo {author} {\bibfnamefont
  {S.}~\bibnamefont {Eaton}},\ }\bibfield  {title} {\bibinfo {title} {Interplay
  of cell dynamics and epithelial tension during morphogenesis of the
  {{Drosophila}} pupal wing},\ }\href {https://doi.org/10.7554/eLife.07090}
  {\bibfield  {journal} {\bibinfo  {journal} {eLife}\ }\textbf {\bibinfo
  {volume} {4}},\ \bibinfo {pages} {e07090} (\bibinfo {year}
  {2015})}\BibitemShut {NoStop}%
\bibitem [{\citenamefont {Forgacs}\ \emph {et~al.}(1998)\citenamefont
  {Forgacs}, \citenamefont {Foty}, \citenamefont {Shafrir},\ and\ \citenamefont
  {Steinberg}}]{forgacs1998viscoelastic}%
  \BibitemOpen
  \bibfield  {author} {\bibinfo {author} {\bibfnamefont {G.}~\bibnamefont
  {Forgacs}}, \bibinfo {author} {\bibfnamefont {R.~A.}\ \bibnamefont {Foty}},
  \bibinfo {author} {\bibfnamefont {Y.}~\bibnamefont {Shafrir}},\ and\ \bibinfo
  {author} {\bibfnamefont {M.~S.}\ \bibnamefont {Steinberg}},\ }\bibfield
  {title} {\bibinfo {title} {Viscoelastic {{Properties}} of {{Living Embryonic
  Tissues}}: A {{Quantitative Study}}},\ }\href
  {https://doi.org/10.1016/S00063495(98)779329} {\bibfield  {journal} {\bibinfo
   {journal} {Biophys. J.}\ }\textbf {\bibinfo {volume} {74}},\ \bibinfo
  {pages} {22272234} (\bibinfo {year} {1998})}\BibitemShut {NoStop}%
\bibitem [{\citenamefont {Hoffman}\ \emph {et~al.}(2006)\citenamefont
  {Hoffman}, \citenamefont {Massiera}, \citenamefont {Van~Citters},\ and\
  \citenamefont {Crocker}}]{Hoffman2006}%
  \BibitemOpen
  \bibfield  {author} {\bibinfo {author} {\bibfnamefont {B.~D.}\ \bibnamefont
  {Hoffman}}, \bibinfo {author} {\bibfnamefont {G.}~\bibnamefont {Massiera}},
  \bibinfo {author} {\bibfnamefont {K.~M.}\ \bibnamefont {Van~Citters}},\ and\
  \bibinfo {author} {\bibfnamefont {J.~C.}\ \bibnamefont {Crocker}},\
  }\bibfield  {title} {\bibinfo {title} {The consensus mechanics of cultured
  mammalian cells},\ }\href {https://doi.org/10.1073/pnas.0510348103}
  {\bibfield  {journal} {\bibinfo  {journal} {Proceedings of the National
  Academy of Sciences}\ }\textbf {\bibinfo {volume} {103}},\ \bibinfo {pages}
  {1025910264} (\bibinfo {year} {2006})}\BibitemShut {NoStop}%
\bibitem [{\citenamefont {Khalilgharibi}\ \emph {et~al.}(2019)\citenamefont
  {Khalilgharibi}, \citenamefont {Fouchard}, \citenamefont {Asadipour},
  \citenamefont {Barrientos}, \citenamefont {Duda}, \citenamefont {Bonfanti},
  \citenamefont {Yonis}, \citenamefont {Harris}, \citenamefont {Mosaffa},
  \citenamefont {Fujita} \emph {et~al.}}]{khalilgharibi2019stress}%
  \BibitemOpen
  \bibfield  {author} {\bibinfo {author} {\bibfnamefont {N.}~\bibnamefont
  {Khalilgharibi}}, \bibinfo {author} {\bibfnamefont {J.}~\bibnamefont
  {Fouchard}}, \bibinfo {author} {\bibfnamefont {N.}~\bibnamefont {Asadipour}},
  \bibinfo {author} {\bibfnamefont {R.}~\bibnamefont {Barrientos}}, \bibinfo
  {author} {\bibfnamefont {M.}~\bibnamefont {Duda}}, \bibinfo {author}
  {\bibfnamefont {A.}~\bibnamefont {Bonfanti}}, \bibinfo {author}
  {\bibfnamefont {A.}~\bibnamefont {Yonis}}, \bibinfo {author} {\bibfnamefont
  {A.}~\bibnamefont {Harris}}, \bibinfo {author} {\bibfnamefont
  {P.}~\bibnamefont {Mosaffa}}, \bibinfo {author} {\bibfnamefont
  {Y.}~\bibnamefont {Fujita}}, \emph {et~al.},\ }\bibfield  {title} {\bibinfo
  {title} {Stress relaxation in epithelial monolayers is controlled by the
  actomyosin cortex},\ }\href@noop {} {\bibfield  {journal} {\bibinfo
  {journal} {Nature physics}\ }\textbf {\bibinfo {volume} {15}},\ \bibinfo
  {pages} {839847} (\bibinfo {year} {2019})}\BibitemShut {NoStop}%
\bibitem [{\citenamefont {Sadeghipour}\ \emph {et~al.}(2018)\citenamefont
  {Sadeghipour}, \citenamefont {Garcia}, \citenamefont {Nelson},\ and\
  \citenamefont {Pruitt}}]{Pruitt2018}%
  \BibitemOpen
  \bibfield  {author} {\bibinfo {author} {\bibfnamefont {E.}~\bibnamefont
  {Sadeghipour}}, \bibinfo {author} {\bibfnamefont {M.~A.}\ \bibnamefont
  {Garcia}}, \bibinfo {author} {\bibfnamefont {W.~J.}\ \bibnamefont {Nelson}},\
  and\ \bibinfo {author} {\bibfnamefont {B.~L.}\ \bibnamefont {Pruitt}},\
  }\bibfield  {title} {\bibinfo {title} {Shearinduced damped oscillations in an
  epithelium depend on actomyosin contraction and ecadherin cell adhesion},\
  }\href {https://doi.org/10.7554/eLife.39640} {\bibfield  {journal} {\bibinfo
  {journal} {eLife}\ }\textbf {\bibinfo {volume} {7}},\ \bibinfo {pages}
  {e39640} (\bibinfo {year} {2018})}\BibitemShut {NoStop}%
\bibitem [{\citenamefont {Bi}\ \emph {et~al.}(2015)\citenamefont {Bi},
  \citenamefont {Lopez}, \citenamefont {Schwarz},\ and\ \citenamefont
  {Manning}}]{bi_nphys_2015}%
  \BibitemOpen
  \bibfield  {author} {\bibinfo {author} {\bibfnamefont {D.}~\bibnamefont
  {Bi}}, \bibinfo {author} {\bibfnamefont {J.~H.}\ \bibnamefont {Lopez}},
  \bibinfo {author} {\bibfnamefont {J.~M.}\ \bibnamefont {Schwarz}},\ and\
  \bibinfo {author} {\bibfnamefont {M.~L.}\ \bibnamefont {Manning}},\
  }\bibfield  {title} {\bibinfo {title} {{A densityindependent rigidity
  transition in biological tissues}},\ }\href
  {https://doi.org/10.1038/nphys3471} {\bibfield  {journal} {\bibinfo
  {journal} {Nature Physics}\ }\textbf {\bibinfo {volume} {11}},\ \bibinfo
  {pages} {10741079} (\bibinfo {year} {2015})}\BibitemShut {NoStop}%
\bibitem [{\citenamefont {Park}\ \emph {et~al.}(2015)\citenamefont {Park},
  \citenamefont {Kim}, \citenamefont {Bi}, \citenamefont {Mitchel},
  \citenamefont {Qazvini}, \citenamefont {Tantisira}, \citenamefont {Park},
  \citenamefont {McGill}, \citenamefont {Kim}, \citenamefont {Gweon},
  \citenamefont {Notbohm}, \citenamefont {Steward}, \citenamefont {Burger},
  \citenamefont {Randell}, \citenamefont {Kho}, \citenamefont {Tambe},
  \citenamefont {Hardin}, \citenamefont {Shore}, \citenamefont {Israel},
  \citenamefont {Weitz}, \citenamefont {Tschumperlin}, \citenamefont {Henske},
  \citenamefont {Weiss}, \citenamefont {Lisa~Manning}, \citenamefont {Butler},
  \citenamefont {Drazen},\ and\ \citenamefont {Fredberg}}]{Park_NMAT_2015}%
  \BibitemOpen
  \bibfield  {author} {\bibinfo {author} {\bibfnamefont {J.}~\bibnamefont
  {Park}}, \bibinfo {author} {\bibfnamefont {J.~H.}\ \bibnamefont {Kim}},
  \bibinfo {author} {\bibfnamefont {D.}~\bibnamefont {Bi}}, \bibinfo {author}
  {\bibfnamefont {J.~A.}\ \bibnamefont {Mitchel}}, \bibinfo {author}
  {\bibfnamefont {N.~T.}\ \bibnamefont {Qazvini}}, \bibinfo {author}
  {\bibfnamefont {K.}~\bibnamefont {Tantisira}}, \bibinfo {author}
  {\bibfnamefont {C.~Y.}\ \bibnamefont {Park}}, \bibinfo {author}
  {\bibfnamefont {M.}~\bibnamefont {McGill}}, \bibinfo {author} {\bibfnamefont
  {S.}~\bibnamefont {Kim}}, \bibinfo {author} {\bibfnamefont {B.}~\bibnamefont
  {Gweon}}, \bibinfo {author} {\bibfnamefont {J.}~\bibnamefont {Notbohm}},
  \bibinfo {author} {\bibfnamefont {R.}~\bibnamefont {Steward}}, \bibinfo
  {author} {\bibfnamefont {S.}~\bibnamefont {Burger}}, \bibinfo {author}
  {\bibfnamefont {S.~H.}\ \bibnamefont {Randell}}, \bibinfo {author}
  {\bibfnamefont {A.~T.}\ \bibnamefont {Kho}}, \bibinfo {author} {\bibfnamefont
  {D.~T.}\ \bibnamefont {Tambe}}, \bibinfo {author} {\bibfnamefont
  {C.}~\bibnamefont {Hardin}}, \bibinfo {author} {\bibfnamefont {S.~A.}\
  \bibnamefont {Shore}}, \bibinfo {author} {\bibfnamefont {E.}~\bibnamefont
  {Israel}}, \bibinfo {author} {\bibfnamefont {D.~A.}\ \bibnamefont {Weitz}},
  \bibinfo {author} {\bibfnamefont {D.~J.}\ \bibnamefont {Tschumperlin}},
  \bibinfo {author} {\bibfnamefont {E.~P.}\ \bibnamefont {Henske}}, \bibinfo
  {author} {\bibfnamefont {S.~T.}\ \bibnamefont {Weiss}}, \bibinfo {author}
  {\bibfnamefont {M.}~\bibnamefont {Lisa~Manning}}, \bibinfo {author}
  {\bibfnamefont {J.~P.}\ \bibnamefont {Butler}}, \bibinfo {author}
  {\bibfnamefont {J.~M.}\ \bibnamefont {Drazen}},\ and\ \bibinfo {author}
  {\bibfnamefont {J.~J.}\ \bibnamefont {Fredberg}},\ }\bibfield  {title}
  {\bibinfo {title} {Unjamming and cell shape in the asthmatic airway
  epithelium},\ }\href {https://doi.org/10.1038/nmat4357} {\bibfield  {journal}
  {\bibinfo  {journal} {Nat Mater}\ }\textbf {\bibinfo {volume} {14}},\
  \bibinfo {pages} {10401048} (\bibinfo {year} {2015})}\BibitemShut {NoStop}%
\bibitem [{\citenamefont {Bi}\ \emph {et~al.}(2016)\citenamefont {Bi},
  \citenamefont {Yang}, \citenamefont {Marchetti},\ and\ \citenamefont
  {Manning}}]{bi2016motility}%
  \BibitemOpen
  \bibfield  {author} {\bibinfo {author} {\bibfnamefont {D.}~\bibnamefont
  {Bi}}, \bibinfo {author} {\bibfnamefont {X.}~\bibnamefont {Yang}}, \bibinfo
  {author} {\bibfnamefont {M.~C.}\ \bibnamefont {Marchetti}},\ and\ \bibinfo
  {author} {\bibfnamefont {M.~L.}\ \bibnamefont {Manning}},\ }\bibfield
  {title} {\bibinfo {title} {Motility-driven glass and jamming transitions in
  biological tissues},\ }\href {https://doi.org/10.1103/PhysRevX.6.021011}
  {\bibfield  {journal} {\bibinfo  {journal} {Phys. Rev. X}\ }\textbf {\bibinfo
  {volume} {6}},\ \bibinfo {pages} {021011} (\bibinfo {year}
  {2016})}\BibitemShut {NoStop}%
\bibitem [{\citenamefont {Malinverno}\ \emph {et~al.}(2017)\citenamefont
  {Malinverno}, \citenamefont {Corallino}, \citenamefont {Giavazzi},
  \citenamefont {Bergert}, \citenamefont {Li}, \citenamefont {Leoni},
  \citenamefont {Disanza}, \citenamefont {Frittoli}, \citenamefont {Oldani},
  \citenamefont {Martini} \emph {et~al.}}]{malinverno2017endocytic}%
  \BibitemOpen
  \bibfield  {author} {\bibinfo {author} {\bibfnamefont {C.}~\bibnamefont
  {Malinverno}}, \bibinfo {author} {\bibfnamefont {S.}~\bibnamefont
  {Corallino}}, \bibinfo {author} {\bibfnamefont {F.}~\bibnamefont {Giavazzi}},
  \bibinfo {author} {\bibfnamefont {M.}~\bibnamefont {Bergert}}, \bibinfo
  {author} {\bibfnamefont {Q.}~\bibnamefont {Li}}, \bibinfo {author}
  {\bibfnamefont {M.}~\bibnamefont {Leoni}}, \bibinfo {author} {\bibfnamefont
  {A.}~\bibnamefont {Disanza}}, \bibinfo {author} {\bibfnamefont
  {E.}~\bibnamefont {Frittoli}}, \bibinfo {author} {\bibfnamefont
  {A.}~\bibnamefont {Oldani}}, \bibinfo {author} {\bibfnamefont
  {E.}~\bibnamefont {Martini}}, \emph {et~al.},\ }\bibfield  {title} {\bibinfo
  {title} {Endocytic reawakening of motility in jammed epithelia},\ }\href@noop
  {} {\bibfield  {journal} {\bibinfo  {journal} {Nature materials}\ }\textbf
  {\bibinfo {volume} {16}},\ \bibinfo {pages} {587} (\bibinfo {year}
  {2017})}\BibitemShut {NoStop}%
\bibitem [{\citenamefont {Mongera}\ \emph {et~al.}(2018)\citenamefont
  {Mongera}, \citenamefont {Rowghanian}, \citenamefont {Gustafson},
  \citenamefont {Shelton}, \citenamefont {Kealhofer}, \citenamefont {Carn},
  \citenamefont {Serwane}, \citenamefont {Lucio}, \citenamefont {Giammona},\
  and\ \citenamefont {Camp{\`a}s}}]{mongera2018fluid}%
  \BibitemOpen
  \bibfield  {author} {\bibinfo {author} {\bibfnamefont {A.}~\bibnamefont
  {Mongera}}, \bibinfo {author} {\bibfnamefont {P.}~\bibnamefont {Rowghanian}},
  \bibinfo {author} {\bibfnamefont {H.~J.}\ \bibnamefont {Gustafson}}, \bibinfo
  {author} {\bibfnamefont {E.}~\bibnamefont {Shelton}}, \bibinfo {author}
  {\bibfnamefont {D.~A.}\ \bibnamefont {Kealhofer}}, \bibinfo {author}
  {\bibfnamefont {E.~K.}\ \bibnamefont {Carn}}, \bibinfo {author}
  {\bibfnamefont {F.}~\bibnamefont {Serwane}}, \bibinfo {author} {\bibfnamefont
  {A.~A.}\ \bibnamefont {Lucio}}, \bibinfo {author} {\bibfnamefont
  {J.}~\bibnamefont {Giammona}},\ and\ \bibinfo {author} {\bibfnamefont
  {O.}~\bibnamefont {Camp{\`a}s}},\ }\bibfield  {title} {\bibinfo {title} {A
  fluidtosolid jamming transition underlies vertebrate body axis elongation},\
  }\href@noop {} {\bibfield  {journal} {\bibinfo  {journal} {Nature}\ }\textbf
  {\bibinfo {volume} {561}},\ \bibinfo {pages} {401} (\bibinfo {year}
  {2018})}\BibitemShut {NoStop}%
\bibitem [{\citenamefont {Lawson-Keister}\ and\ \citenamefont
  {Manning}(2021)}]{lawson2021jamming}%
  \BibitemOpen
  \bibfield  {author} {\bibinfo {author} {\bibfnamefont {E.}~\bibnamefont
  {Lawson-Keister}}\ and\ \bibinfo {author} {\bibfnamefont {M.~L.}\
  \bibnamefont {Manning}},\ }\bibfield  {title} {\bibinfo {title} {Jamming and
  arrest of cell motion in biological tissues},\ }\href@noop {} {\bibfield
  {journal} {\bibinfo  {journal} {Current Opinion in Cell Biology}\ }\textbf
  {\bibinfo {volume} {72}},\ \bibinfo {pages} {146} (\bibinfo {year}
  {2021})}\BibitemShut {NoStop}%
\bibitem [{\citenamefont {Fern{\'a}ndez}\ \emph {et~al.}(2006)\citenamefont
  {Fern{\'a}ndez}, \citenamefont {Pullarkat},\ and\ \citenamefont
  {Ott}}]{fernandez2006master}%
  \BibitemOpen
  \bibfield  {author} {\bibinfo {author} {\bibfnamefont {P.}~\bibnamefont
  {Fern{\'a}ndez}}, \bibinfo {author} {\bibfnamefont {P.~A.}\ \bibnamefont
  {Pullarkat}},\ and\ \bibinfo {author} {\bibfnamefont {A.}~\bibnamefont
  {Ott}},\ }\bibfield  {title} {\bibinfo {title} {A master relation defines the
  nonlinear viscoelasticity of single fibroblasts},\ }\href@noop {} {\bibfield
  {journal} {\bibinfo  {journal} {Biophysical journal}\ }\textbf {\bibinfo
  {volume} {90}},\ \bibinfo {pages} {37963805} (\bibinfo {year}
  {2006})}\BibitemShut {NoStop}%
\bibitem [{\citenamefont {Trepat}\ \emph {et~al.}(2007)\citenamefont {Trepat},
  \citenamefont {Deng}, \citenamefont {An}, \citenamefont {Navajas},
  \citenamefont {Tschumperlin}, \citenamefont {Gerthoffer}, \citenamefont
  {Butler},\ and\ \citenamefont
  {Fredberg}}]{trepat_fredberg_stretch_nature_2007}%
  \BibitemOpen
  \bibfield  {author} {\bibinfo {author} {\bibfnamefont {X.}~\bibnamefont
  {Trepat}}, \bibinfo {author} {\bibfnamefont {L.}~\bibnamefont {Deng}},
  \bibinfo {author} {\bibfnamefont {S.~S.}\ \bibnamefont {An}}, \bibinfo
  {author} {\bibfnamefont {D.}~\bibnamefont {Navajas}}, \bibinfo {author}
  {\bibfnamefont {D.~J.}\ \bibnamefont {Tschumperlin}}, \bibinfo {author}
  {\bibfnamefont {W.~T.}\ \bibnamefont {Gerthoffer}}, \bibinfo {author}
  {\bibfnamefont {J.~P.}\ \bibnamefont {Butler}},\ and\ \bibinfo {author}
  {\bibfnamefont {J.~J.}\ \bibnamefont {Fredberg}},\ }\bibfield  {title}
  {\bibinfo {title} {Universal physical responses to stretch in the living
  cell},\ }\href@noop {} {\bibfield  {journal} {\bibinfo  {journal} {Nature}\
  }\textbf {\bibinfo {volume} {447}},\ \bibinfo {pages} {592595} (\bibinfo
  {year} {2007})}\BibitemShut {NoStop}%
\bibitem [{\citenamefont {Harris}\ \emph {et~al.}(2012)\citenamefont {Harris},
  \citenamefont {Peter}, \citenamefont {Bellis}, \citenamefont {Baum},
  \citenamefont {Kabla},\ and\ \citenamefont {Charras}}]{Harris_PNAS_stretch}%
  \BibitemOpen
  \bibfield  {author} {\bibinfo {author} {\bibfnamefont {A.~R.}\ \bibnamefont
  {Harris}}, \bibinfo {author} {\bibfnamefont {L.}~\bibnamefont {Peter}},
  \bibinfo {author} {\bibfnamefont {J.}~\bibnamefont {Bellis}}, \bibinfo
  {author} {\bibfnamefont {B.}~\bibnamefont {Baum}}, \bibinfo {author}
  {\bibfnamefont {A.~J.}\ \bibnamefont {Kabla}},\ and\ \bibinfo {author}
  {\bibfnamefont {G.~T.}\ \bibnamefont {Charras}},\ }\bibfield  {title}
  {\bibinfo {title} {Characterizing the mechanics of cultured cell
  monolayers},\ }\href {https://doi.org/10.1073/pnas.1213301109} {\bibfield
  {journal} {\bibinfo  {journal} {Proceedings of the National Academy of
  Sciences}\ }\textbf {\bibinfo {volume} {109}},\ \bibinfo {pages} {1644916454}
  (\bibinfo {year} {2012})}\BibitemShut {NoStop}%
\bibitem [{\citenamefont {Latorre}\ \emph {et~al.}(2018)\citenamefont
  {Latorre}, \citenamefont {Kale}, \citenamefont {Casares}, \citenamefont
  {G{\'o}mezGonz{\'a}lez}, \citenamefont {Uroz}, \citenamefont {Valon},
  \citenamefont {Nair}, \citenamefont {Garreta}, \citenamefont {Montserrat},
  \citenamefont {Del~Campo} \emph {et~al.}}]{latorre2018active}%
  \BibitemOpen
  \bibfield  {author} {\bibinfo {author} {\bibfnamefont {E.}~\bibnamefont
  {Latorre}}, \bibinfo {author} {\bibfnamefont {S.}~\bibnamefont {Kale}},
  \bibinfo {author} {\bibfnamefont {L.}~\bibnamefont {Casares}}, \bibinfo
  {author} {\bibfnamefont {M.}~\bibnamefont {G{\'o}mezGonz{\'a}lez}}, \bibinfo
  {author} {\bibfnamefont {M.}~\bibnamefont {Uroz}}, \bibinfo {author}
  {\bibfnamefont {L.}~\bibnamefont {Valon}}, \bibinfo {author} {\bibfnamefont
  {R.~V.}\ \bibnamefont {Nair}}, \bibinfo {author} {\bibfnamefont
  {E.}~\bibnamefont {Garreta}}, \bibinfo {author} {\bibfnamefont
  {N.}~\bibnamefont {Montserrat}}, \bibinfo {author} {\bibfnamefont
  {A.}~\bibnamefont {Del~Campo}}, \emph {et~al.},\ }\bibfield  {title}
  {\bibinfo {title} {Active superelasticity in threedimensional epithelia of
  controlled shape},\ }\href@noop {} {\bibfield  {journal} {\bibinfo  {journal}
  {Nature}\ }\textbf {\bibinfo {volume} {563}},\ \bibinfo {pages} {203208}
  (\bibinfo {year} {2018})}\BibitemShut {NoStop}%
\bibitem [{\citenamefont {Prakash}\ \emph {et~al.}(2021)\citenamefont
  {Prakash}, \citenamefont {Bull},\ and\ \citenamefont
  {Prakash}}]{prakash2021motility}%
  \BibitemOpen
  \bibfield  {author} {\bibinfo {author} {\bibfnamefont {V.~N.}\ \bibnamefont
  {Prakash}}, \bibinfo {author} {\bibfnamefont {M.~S.}\ \bibnamefont {Bull}},\
  and\ \bibinfo {author} {\bibfnamefont {M.}~\bibnamefont {Prakash}},\
  }\bibfield  {title} {\bibinfo {title} {Motilityinduced fracture reveals a
  ductiletobrittle crossover in a simple animal’s epithelia},\ }\href@noop {}
  {\bibfield  {journal} {\bibinfo  {journal} {Nature Physics}\ }\textbf
  {\bibinfo {volume} {17}},\ \bibinfo {pages} {504511} (\bibinfo {year}
  {2021})}\BibitemShut {NoStop}%
\bibitem [{\citenamefont {Nagai}\ and\ \citenamefont
  {Honda}(2001)}]{Nagai_PMB_2001}%
  \BibitemOpen
  \bibfield  {author} {\bibinfo {author} {\bibfnamefont {T.}~\bibnamefont
  {Nagai}}\ and\ \bibinfo {author} {\bibfnamefont {H.}~\bibnamefont {Honda}},\
  }\bibfield  {title} {\bibinfo {title} {A dynamic cell model for the formation
  of epithelial tissues},\ }\href {https://doi.org/10.1080/13642810108205772}
  {\bibfield  {journal} {\bibinfo  {journal} {Philosophical Magazine B}\
  }\textbf {\bibinfo {volume} {81}},\ \bibinfo {pages} {699719} (\bibinfo
  {year} {2001})}\BibitemShut {NoStop}%
\bibitem [{\citenamefont {Farhadifar}\ \emph {et~al.}(2007)\citenamefont
  {Farhadifar}, \citenamefont {Röper}, \citenamefont {Aigouy}, \citenamefont
  {Eaton},\ and\ \citenamefont {Jülicher}}]{Farhadifar_CB_2007}%
  \BibitemOpen
  \bibfield  {author} {\bibinfo {author} {\bibfnamefont {R.}~\bibnamefont
  {Farhadifar}}, \bibinfo {author} {\bibfnamefont {J.}~\bibnamefont {Röper}},
  \bibinfo {author} {\bibfnamefont {B.}~\bibnamefont {Aigouy}}, \bibinfo
  {author} {\bibfnamefont {S.}~\bibnamefont {Eaton}},\ and\ \bibinfo {author}
  {\bibfnamefont {F.}~\bibnamefont {Jülicher}},\ }\bibfield  {title} {\bibinfo
  {title} {The influence of cell mechanics, cellcell interactions, and
  proliferation on epithelial packing},\ }\href
  {https://doi.org/https://doi.org/10.1016/j.cub.2007.11.049} {\bibfield
  {journal} {\bibinfo  {journal} {Current Biology}\ }\textbf {\bibinfo {volume}
  {17}},\ \bibinfo {pages} {2095 2104} (\bibinfo {year} {2007})}\BibitemShut
  {NoStop}%
\bibitem [{\citenamefont {Moshe}\ \emph {et~al.}(2018)\citenamefont {Moshe},
  \citenamefont {Bowick},\ and\ \citenamefont
  {Marchetti}}]{moshe2018geometric}%
  \BibitemOpen
  \bibfield  {author} {\bibinfo {author} {\bibfnamefont {M.}~\bibnamefont
  {Moshe}}, \bibinfo {author} {\bibfnamefont {M.~J.}\ \bibnamefont {Bowick}},\
  and\ \bibinfo {author} {\bibfnamefont {M.~C.}\ \bibnamefont {Marchetti}},\
  }\bibfield  {title} {\bibinfo {title} {Geometric frustration and solid-solid
  transitions in model 2d tissue},\ }\href@noop {} {\bibfield  {journal}
  {\bibinfo  {journal} {Physical review letters}\ }\textbf {\bibinfo {volume}
  {120}},\ \bibinfo {pages} {268105} (\bibinfo {year} {2018})}\BibitemShut
  {NoStop}%
\bibitem [{\citenamefont {Tong}\ \emph {et~al.}(2021)\citenamefont {Tong},
  \citenamefont {Singh}, \citenamefont {Sknepnek},\ and\ \citenamefont
  {Kosmrlj}}]{Rastko_2021}%
  \BibitemOpen
  \bibfield  {author} {\bibinfo {author} {\bibfnamefont {S.}~\bibnamefont
  {Tong}}, \bibinfo {author} {\bibfnamefont {N.~K.}\ \bibnamefont {Singh}},
  \bibinfo {author} {\bibfnamefont {R.}~\bibnamefont {Sknepnek}},\ and\
  \bibinfo {author} {\bibfnamefont {A.}~\bibnamefont {Kosmrlj}},\ }\href@noop
  {} {\bibinfo {title} {Linear viscoelastic properties of the vertex model for
  epithelial tissues}} (\bibinfo {year} {2021}),\ \Eprint
  {https://arxiv.org/abs/2102.11181} {arXiv:2102.11181 [condmat.soft]}
  \BibitemShut {NoStop}%
\bibitem [{\citenamefont {Hernandez}\ \emph {et~al.}(2022)\citenamefont
  {Hernandez}, \citenamefont {Staddon}, \citenamefont {Bowick}, \citenamefont
  {Marchetti},\ and\ \citenamefont {Moshe}}]{hernandez2022anomalous}%
  \BibitemOpen
  \bibfield  {author} {\bibinfo {author} {\bibfnamefont {A.}~\bibnamefont
  {Hernandez}}, \bibinfo {author} {\bibfnamefont {M.~F.}\ \bibnamefont
  {Staddon}}, \bibinfo {author} {\bibfnamefont {M.~J.}\ \bibnamefont {Bowick}},
  \bibinfo {author} {\bibfnamefont {M.~C.}\ \bibnamefont {Marchetti}},\ and\
  \bibinfo {author} {\bibfnamefont {M.}~\bibnamefont {Moshe}},\ }\bibfield
  {title} {\bibinfo {title} {Anomalous elasticity of a cellular tissue vertex
  model},\ }\href@noop {} {\bibfield  {journal} {\bibinfo  {journal} {Physical
  Review E}\ }\textbf {\bibinfo {volume} {105}},\ \bibinfo {pages} {064611}
  (\bibinfo {year} {2022})}\BibitemShut {NoStop}%
\bibitem [{\citenamefont {Merzouki}\ \emph {et~al.}(2016)\citenamefont
  {Merzouki}, \citenamefont {Malaspinas},\ and\ \citenamefont
  {Chopard}}]{Merzouki_vm_strain}%
  \BibitemOpen
  \bibfield  {author} {\bibinfo {author} {\bibfnamefont {A.}~\bibnamefont
  {Merzouki}}, \bibinfo {author} {\bibfnamefont {O.}~\bibnamefont
  {Malaspinas}},\ and\ \bibinfo {author} {\bibfnamefont {B.}~\bibnamefont
  {Chopard}},\ }\bibfield  {title} {\bibinfo {title} {The mechanical properties
  of a cellbased numerical model of epithelium},\ }\href
  {https://doi.org/10.1039/C6SM00106H} {\bibfield  {journal} {\bibinfo
  {journal} {Soft Matter}\ }\textbf {\bibinfo {volume} {12}},\ \bibinfo {pages}
  {47454754} (\bibinfo {year} {2016})}\BibitemShut {NoStop}%
\bibitem [{\citenamefont {Popovi{\'{c}}}\ \emph {et~al.}(2021)\citenamefont
  {Popovi{\'{c}}}, \citenamefont {Druelle}, \citenamefont {Dye}, \citenamefont
  {Jülicher},\ and\ \citenamefont {Wyart}}]{Popovic_2021}%
  \BibitemOpen
  \bibfield  {author} {\bibinfo {author} {\bibfnamefont {M.}~\bibnamefont
  {Popovi{\'{c}}}}, \bibinfo {author} {\bibfnamefont {V.}~\bibnamefont
  {Druelle}}, \bibinfo {author} {\bibfnamefont {N.~A.}\ \bibnamefont {Dye}},
  \bibinfo {author} {\bibfnamefont {F.}~\bibnamefont {Jülicher}},\ and\
  \bibinfo {author} {\bibfnamefont {M.}~\bibnamefont {Wyart}},\ }\bibfield
  {title} {\bibinfo {title} {Inferring the flow properties of epithelial
  tissues from their geometry},\ }\href
  {https://doi.org/10.1088/13672630/abcbc7} {\bibfield  {journal} {\bibinfo
  {journal} {New Journal of Physics}\ }\textbf {\bibinfo {volume} {23}},\
  \bibinfo {pages} {033004} (\bibinfo {year} {2021})}\BibitemShut {NoStop}%
\bibitem [{\citenamefont {Duclut}\ \emph {et~al.}(2021)\citenamefont {Duclut},
  \citenamefont {Paijmans}, \citenamefont {Inamdar}, \citenamefont {Modes},\
  and\ \citenamefont {J{\"u}licher}}]{duclut2021nonlinear}%
  \BibitemOpen
  \bibfield  {author} {\bibinfo {author} {\bibfnamefont {C.}~\bibnamefont
  {Duclut}}, \bibinfo {author} {\bibfnamefont {J.}~\bibnamefont {Paijmans}},
  \bibinfo {author} {\bibfnamefont {M.~M.}\ \bibnamefont {Inamdar}}, \bibinfo
  {author} {\bibfnamefont {C.~D.}\ \bibnamefont {Modes}},\ and\ \bibinfo
  {author} {\bibfnamefont {F.}~\bibnamefont {J{\"u}licher}},\ }\bibfield
  {title} {\bibinfo {title} {Nonlinear rheology of cellular networks},\
  }\href@noop {} {\bibfield  {journal} {\bibinfo  {journal} {Cells \&
  development}\ }\textbf {\bibinfo {volume} {168}},\ \bibinfo {pages} {203746}
  (\bibinfo {year} {2021})}\BibitemShut {NoStop}%
\bibitem [{\citenamefont {Pasupalak}\ \emph {et~al.}(2021)\citenamefont
  {Pasupalak}, \citenamefont {Samidurai}, \citenamefont {Li}, \citenamefont
  {Zheng}, \citenamefont {Ni},\ and\ \citenamefont
  {Ciamarra}}]{PicaCiamarra_rheology}%
  \BibitemOpen
  \bibfield  {author} {\bibinfo {author} {\bibfnamefont {A.}~\bibnamefont
  {Pasupalak}}, \bibinfo {author} {\bibfnamefont {S.~K.}\ \bibnamefont
  {Samidurai}}, \bibinfo {author} {\bibfnamefont {Y.}~\bibnamefont {Li}},
  \bibinfo {author} {\bibfnamefont {Y.}~\bibnamefont {Zheng}}, \bibinfo
  {author} {\bibfnamefont {R.}~\bibnamefont {Ni}},\ and\ \bibinfo {author}
  {\bibfnamefont {M.~P.}\ \bibnamefont {Ciamarra}},\ }\bibfield  {title}
  {\bibinfo {title} {Unconventional rheological properties in systems of
  deformable particles},\ }\href {https://doi.org/10.1039/D1SM00936B}
  {\bibfield  {journal} {\bibinfo  {journal} {Soft Matter}\ }\textbf {\bibinfo
  {volume} {17}},\ \bibinfo {pages} {77087713} (\bibinfo {year}
  {2021})}\BibitemShut {NoStop}%
\bibitem [{\citenamefont {Huang}\ \emph {et~al.}(2022)\citenamefont {Huang},
  \citenamefont {Cochran}, \citenamefont {Fielding}, \citenamefont
  {Marchetti},\ and\ \citenamefont {Bi}}]{huang2022shear}%
  \BibitemOpen
  \bibfield  {author} {\bibinfo {author} {\bibfnamefont {J.}~\bibnamefont
  {Huang}}, \bibinfo {author} {\bibfnamefont {J.~O.}\ \bibnamefont {Cochran}},
  \bibinfo {author} {\bibfnamefont {S.~M.}\ \bibnamefont {Fielding}}, \bibinfo
  {author} {\bibfnamefont {M.~C.}\ \bibnamefont {Marchetti}},\ and\ \bibinfo
  {author} {\bibfnamefont {D.}~\bibnamefont {Bi}},\ }\bibfield  {title}
  {\bibinfo {title} {Sheardriven solidification and nonlinear elasticity in
  epithelial tissues},\ }\href@noop {} {\bibfield  {journal} {\bibinfo
  {journal} {Physical Review Letters}\ }\textbf {\bibinfo {volume} {128}},\
  \bibinfo {pages} {178001} (\bibinfo {year} {2022})}\BibitemShut {NoStop}%
\bibitem [{\citenamefont {Czajkowski}\ \emph {et~al.}(2018)\citenamefont
  {Czajkowski}, \citenamefont {Bi}, \citenamefont {Manning},\ and\
  \citenamefont {Marchetti}}]{czajkowski2018hydrodynamics}%
  \BibitemOpen
  \bibfield  {author} {\bibinfo {author} {\bibfnamefont {M.}~\bibnamefont
  {Czajkowski}}, \bibinfo {author} {\bibfnamefont {D.}~\bibnamefont {Bi}},
  \bibinfo {author} {\bibfnamefont {M.~L.}\ \bibnamefont {Manning}},\ and\
  \bibinfo {author} {\bibfnamefont {M.~C.}\ \bibnamefont {Marchetti}},\
  }\bibfield  {title} {\bibinfo {title} {Hydrodynamics of shape-driven rigidity
  transitions in motile tissues},\ }\href@noop {} {\bibfield  {journal}
  {\bibinfo  {journal} {Soft Matter}\ }\textbf {\bibinfo {volume} {14}},\
  \bibinfo {pages} {5628} (\bibinfo {year} {2018})}\BibitemShut {NoStop}%
\bibitem [{\citenamefont {Marchetti}\ \emph {et~al.}(2013)\citenamefont
  {Marchetti}, \citenamefont {Joanny}, \citenamefont {Ramaswamy}, \citenamefont
  {Liverpool}, \citenamefont {Prost}, \citenamefont {Rao},\ and\ \citenamefont
  {Simha}}]{marchetti2013hydrodynamics}%
  \BibitemOpen
  \bibfield  {author} {\bibinfo {author} {\bibfnamefont {M.~C.}\ \bibnamefont
  {Marchetti}}, \bibinfo {author} {\bibfnamefont {J.~F.}\ \bibnamefont
  {Joanny}}, \bibinfo {author} {\bibfnamefont {S.}~\bibnamefont {Ramaswamy}},
  \bibinfo {author} {\bibfnamefont {T.~B.}\ \bibnamefont {Liverpool}}, \bibinfo
  {author} {\bibfnamefont {J.}~\bibnamefont {Prost}}, \bibinfo {author}
  {\bibfnamefont {M.}~\bibnamefont {Rao}},\ and\ \bibinfo {author}
  {\bibfnamefont {R.~A.}\ \bibnamefont {Simha}},\ }\bibfield  {title} {\bibinfo
  {title} {Hydrodynamics of soft active matter},\ }\href
  {https://doi.org/10.1103/RevModPhys.85.1143} {\bibfield  {journal} {\bibinfo
  {journal} {Rev. Mod. Phys.}\ }\textbf {\bibinfo {volume} {85}},\ \bibinfo
  {pages} {11431189} (\bibinfo {year} {2013})}\BibitemShut {NoStop}%
\bibitem [{\citenamefont {Prost}\ \emph {et~al.}(2015)\citenamefont {Prost},
  \citenamefont {J{\"u}licher},\ and\ \citenamefont
  {Joanny}}]{prost2015active}%
  \BibitemOpen
  \bibfield  {author} {\bibinfo {author} {\bibfnamefont {J.}~\bibnamefont
  {Prost}}, \bibinfo {author} {\bibfnamefont {F.}~\bibnamefont
  {J{\"u}licher}},\ and\ \bibinfo {author} {\bibfnamefont {J.}~\bibnamefont
  {Joanny}},\ }\bibfield  {title} {\bibinfo {title} {Active gel physics},\
  }\href {https://doi.org/10.1038/nphys3224} {\bibfield  {journal} {\bibinfo
  {journal} {Nat. Phys.}\ }\textbf {\bibinfo {volume} {11}},\ \bibinfo {pages}
  {111117} (\bibinfo {year} {2015})}\BibitemShut {NoStop}%
\bibitem [{\citenamefont {Ranft}\ \emph {et~al.}(2010)\citenamefont {Ranft},
  \citenamefont {Basan}, \citenamefont {Elgeti}, \citenamefont {Joanny},
  \citenamefont {Prost},\ and\ \citenamefont
  {J{\"u}licher}}]{ranft2010fluidization}%
  \BibitemOpen
  \bibfield  {author} {\bibinfo {author} {\bibfnamefont {J.}~\bibnamefont
  {Ranft}}, \bibinfo {author} {\bibfnamefont {M.}~\bibnamefont {Basan}},
  \bibinfo {author} {\bibfnamefont {J.}~\bibnamefont {Elgeti}}, \bibinfo
  {author} {\bibfnamefont {J.}~\bibnamefont {Joanny}}, \bibinfo {author}
  {\bibfnamefont {J.}~\bibnamefont {Prost}},\ and\ \bibinfo {author}
  {\bibfnamefont {F.}~\bibnamefont {J{\"u}licher}},\ }\bibfield  {title}
  {\bibinfo {title} {Fluidization of tissues by cell division and apoptosis},\
  }\href {https://doi.org/10.1073/pnas.1011086107} {\bibfield  {journal}
  {\bibinfo  {journal} {Proceedings of the National Academy of Sciences}\
  }\textbf {\bibinfo {volume} {107}},\ \bibinfo {pages} {2086320868} (\bibinfo
  {year} {2010})}\BibitemShut {NoStop}%
\bibitem [{\citenamefont {Dye}\ \emph {et~al.}(2021)\citenamefont {Dye},
  \citenamefont {Popovi{\'c}}, \citenamefont {Iyer}, \citenamefont {Fuhrmann},
  \citenamefont {PiscitelloG{\'o}mez}, \citenamefont {Eaton},\ and\
  \citenamefont {J{\"u}licher}}]{dye2021self}%
  \BibitemOpen
  \bibfield  {author} {\bibinfo {author} {\bibfnamefont {N.~A.}\ \bibnamefont
  {Dye}}, \bibinfo {author} {\bibfnamefont {M.}~\bibnamefont {Popovi{\'c}}},
  \bibinfo {author} {\bibfnamefont {K.~V.}\ \bibnamefont {Iyer}}, \bibinfo
  {author} {\bibfnamefont {J.~F.}\ \bibnamefont {Fuhrmann}}, \bibinfo {author}
  {\bibfnamefont {R.}~\bibnamefont {PiscitelloG{\'o}mez}}, \bibinfo {author}
  {\bibfnamefont {S.}~\bibnamefont {Eaton}},\ and\ \bibinfo {author}
  {\bibfnamefont {F.}~\bibnamefont {J{\"u}licher}},\ }\bibfield  {title}
  {\bibinfo {title} {Selforganized patterning of cell morphology via
  mechanosensitive feedback},\ }\href@noop {} {\bibfield  {journal} {\bibinfo
  {journal} {Elife}\ }\textbf {\bibinfo {volume} {10}},\ \bibinfo {pages}
  {e57964} (\bibinfo {year} {2021})}\BibitemShut {NoStop}%
\bibitem [{\citenamefont {Grossman}\ and\ \citenamefont
  {Joanny}(2022)}]{grossman2022instabilities}%
  \BibitemOpen
  \bibfield  {author} {\bibinfo {author} {\bibfnamefont {D.}~\bibnamefont
  {Grossman}}\ and\ \bibinfo {author} {\bibfnamefont {J.-F.}\ \bibnamefont
  {Joanny}},\ }\bibfield  {title} {\bibinfo {title} {Instabilities and geometry
  of growing tissues},\ }\href@noop {} {\bibfield  {journal} {\bibinfo
  {journal} {Physical Review Letters}\ }\textbf {\bibinfo {volume} {129}},\
  \bibinfo {pages} {048102} (\bibinfo {year} {2022})}\BibitemShut {NoStop}%
\bibitem [{\citenamefont {Ishihara}\ \emph {et~al.}(2017)\citenamefont
  {Ishihara}, \citenamefont {Marcq},\ and\ \citenamefont
  {Sugimura}}]{ishihara2017cells}%
  \BibitemOpen
  \bibfield  {author} {\bibinfo {author} {\bibfnamefont {S.}~\bibnamefont
  {Ishihara}}, \bibinfo {author} {\bibfnamefont {P.}~\bibnamefont {Marcq}},\
  and\ \bibinfo {author} {\bibfnamefont {K.}~\bibnamefont {Sugimura}},\
  }\bibfield  {title} {\bibinfo {title} {From cells to tissue: A continuum
  model of epithelial mechanics},\ }\href@noop {} {\bibfield  {journal}
  {\bibinfo  {journal} {Physical Review E}\ }\textbf {\bibinfo {volume} {96}},\
  \bibinfo {pages} {022418} (\bibinfo {year} {2017})}\BibitemShut {NoStop}%
\bibitem [{\citenamefont {Murisic}\ \emph {et~al.}(2015)\citenamefont
  {Murisic}, \citenamefont {Hakim}, \citenamefont {Kevrekidis}, \citenamefont
  {Shvartsman},\ and\ \citenamefont {Audoly}}]{murisic2015discrete}%
  \BibitemOpen
  \bibfield  {author} {\bibinfo {author} {\bibfnamefont {N.}~\bibnamefont
  {Murisic}}, \bibinfo {author} {\bibfnamefont {V.}~\bibnamefont {Hakim}},
  \bibinfo {author} {\bibfnamefont {I.~G.}\ \bibnamefont {Kevrekidis}},
  \bibinfo {author} {\bibfnamefont {S.~Y.}\ \bibnamefont {Shvartsman}},\ and\
  \bibinfo {author} {\bibfnamefont {B.}~\bibnamefont {Audoly}},\ }\bibfield
  {title} {\bibinfo {title} {From discrete to continuum models of
  three-dimensional deformations in epithelial sheets},\ }\href@noop {}
  {\bibfield  {journal} {\bibinfo  {journal} {Biophysical journal}\ }\textbf
  {\bibinfo {volume} {109}},\ \bibinfo {pages} {154} (\bibinfo {year}
  {2015})}\BibitemShut {NoStop}%
\bibitem [{\citenamefont {Picard}\ \emph {et~al.}(2002)\citenamefont {Picard},
  \citenamefont {Ajdari}, \citenamefont {Bocquet},\ and\ \citenamefont
  {Lequeux}}]{picard2002simple}%
  \BibitemOpen
  \bibfield  {author} {\bibinfo {author} {\bibfnamefont {G.}~\bibnamefont
  {Picard}}, \bibinfo {author} {\bibfnamefont {A.}~\bibnamefont {Ajdari}},
  \bibinfo {author} {\bibfnamefont {L.}~\bibnamefont {Bocquet}},\ and\ \bibinfo
  {author} {\bibfnamefont {F.}~\bibnamefont {Lequeux}},\ }\bibfield  {title}
  {\bibinfo {title} {Simple model for heterogeneous flows of yield stress
  fluids},\ }\href@noop {} {\bibfield  {journal} {\bibinfo  {journal} {Physical
  Review E}\ }\textbf {\bibinfo {volume} {66}},\ \bibinfo {pages} {051501}
  (\bibinfo {year} {2002})}\BibitemShut {NoStop}%
\bibitem [{\citenamefont {Staple}\ \emph {et~al.}(2010)\citenamefont {Staple},
  \citenamefont {Farhadifar}, \citenamefont {R{\"o}per}, \citenamefont
  {Aigouy}, \citenamefont {Eaton},\ and\ \citenamefont
  {J{\"u}licher}}]{Staple_PJE_2010}%
  \BibitemOpen
  \bibfield  {author} {\bibinfo {author} {\bibfnamefont {D.~B.}\ \bibnamefont
  {Staple}}, \bibinfo {author} {\bibfnamefont {R.}~\bibnamefont {Farhadifar}},
  \bibinfo {author} {\bibfnamefont {J.~C.}\ \bibnamefont {R{\"o}per}}, \bibinfo
  {author} {\bibfnamefont {B.}~\bibnamefont {Aigouy}}, \bibinfo {author}
  {\bibfnamefont {S.}~\bibnamefont {Eaton}},\ and\ \bibinfo {author}
  {\bibfnamefont {F.}~\bibnamefont {J{\"u}licher}},\ }\bibfield  {title}
  {\bibinfo {title} {Mechanics and remodelling of cell packings in epithelia},\
  }\href {https://doi.org/10.1140/epje/i2010106770} {\bibfield  {journal}
  {\bibinfo  {journal} {The European Physical Journal E}\ }\textbf {\bibinfo
  {volume} {33}},\ \bibinfo {pages} {117127} (\bibinfo {year}
  {2010})}\BibitemShut {NoStop}%
\bibitem [{\citenamefont {Yang}\ \emph {et~al.}(2017)\citenamefont {Yang},
  \citenamefont {Bi}, \citenamefont {Czajkowski}, \citenamefont {Merkel},
  \citenamefont {Manning},\ and\ \citenamefont
  {Marchetti}}]{yang2017correlating}%
  \BibitemOpen
  \bibfield  {author} {\bibinfo {author} {\bibfnamefont {X.}~\bibnamefont
  {Yang}}, \bibinfo {author} {\bibfnamefont {D.}~\bibnamefont {Bi}}, \bibinfo
  {author} {\bibfnamefont {M.}~\bibnamefont {Czajkowski}}, \bibinfo {author}
  {\bibfnamefont {M.}~\bibnamefont {Merkel}}, \bibinfo {author} {\bibfnamefont
  {M.~L.}\ \bibnamefont {Manning}},\ and\ \bibinfo {author} {\bibfnamefont
  {M.~C.}\ \bibnamefont {Marchetti}},\ }\bibfield  {title} {\bibinfo {title}
  {Correlating cell shape and cellular stress in motile confluent tissues},\
  }\href@noop {} {\bibfield  {journal} {\bibinfo  {journal} {Proceedings of the
  National Academy of Sciences}\ }\textbf {\bibinfo {volume} {114}},\ \bibinfo
  {pages} {12663} (\bibinfo {year} {2017})}\BibitemShut {NoStop}%
\bibitem [{\citenamefont {Das}\ \emph {et~al.}(2021)\citenamefont {Das},
  \citenamefont {Sastry},\ and\ \citenamefont {Bi}}]{Das_T1}%
  \BibitemOpen
  \bibfield  {author} {\bibinfo {author} {\bibfnamefont {A.}~\bibnamefont
  {Das}}, \bibinfo {author} {\bibfnamefont {S.}~\bibnamefont {Sastry}},\ and\
  \bibinfo {author} {\bibfnamefont {D.}~\bibnamefont {Bi}},\ }\bibfield
  {title} {\bibinfo {title} {Controlled neighbor exchanges drive glassy
  behavior, intermittency, and cell streaming in epithelial tissues},\ }\href
  {https://doi.org/10.1103/PhysRevX.11.041037} {\bibfield  {journal} {\bibinfo
  {journal} {Phys. Rev. X}\ }\textbf {\bibinfo {volume} {11}},\ \bibinfo
  {pages} {041037} (\bibinfo {year} {2021})}\BibitemShut {NoStop}%
\bibitem [{our()}]{ourSI}%
  \BibitemOpen
  \href@noop {} {}\bibinfo {note} {See Supplemental Material}\BibitemShut
  {NoStop}%
\bibitem [{\citenamefont {Yan}\ and\ \citenamefont
  {Bi}(2019)}]{yan_bi_rigidity}%
  \BibitemOpen
  \bibfield  {author} {\bibinfo {author} {\bibfnamefont {L.}~\bibnamefont
  {Yan}}\ and\ \bibinfo {author} {\bibfnamefont {D.}~\bibnamefont {Bi}},\
  }\bibfield  {title} {\bibinfo {title} {Multicellular rosettes drive
  fluid-solid transition in epithelial tissues},\ }\href
  {https://doi.org/10.1103/PhysRevX.9.011029} {\bibfield  {journal} {\bibinfo
  {journal} {Phys. Rev. X}\ }\textbf {\bibinfo {volume} {9}},\ \bibinfo {pages}
  {011029} (\bibinfo {year} {2019})}\BibitemShut {NoStop}%
\bibitem [{\citenamefont {Wang}\ \emph {et~al.}(2020)\citenamefont {Wang},
  \citenamefont {Merkel}, \citenamefont {Sutter}, \citenamefont
  {ErdemciTandogan}, \citenamefont {Manning},\ and\ \citenamefont
  {Kasza}}]{wang2020anisotropy}%
  \BibitemOpen
  \bibfield  {author} {\bibinfo {author} {\bibfnamefont {X.}~\bibnamefont
  {Wang}}, \bibinfo {author} {\bibfnamefont {M.}~\bibnamefont {Merkel}},
  \bibinfo {author} {\bibfnamefont {L.~B.}\ \bibnamefont {Sutter}}, \bibinfo
  {author} {\bibfnamefont {G.}~\bibnamefont {ErdemciTandogan}}, \bibinfo
  {author} {\bibfnamefont {M.~L.}\ \bibnamefont {Manning}},\ and\ \bibinfo
  {author} {\bibfnamefont {K.~E.}\ \bibnamefont {Kasza}},\ }\bibfield  {title}
  {\bibinfo {title} {Anisotropy links cell shapes to tissue flow during
  convergent extension},\ }\href@noop {} {\bibfield  {journal} {\bibinfo
  {journal} {Proceedings of the National Academy of Sciences}\ }\textbf
  {\bibinfo {volume} {117}},\ \bibinfo {pages} {1354113551} (\bibinfo {year}
  {2020})}\BibitemShut {NoStop}%
\bibitem [{\citenamefont {Merkel}\ and\ \citenamefont
  {Manning}(2018)}]{merkel2018geometrically}%
  \BibitemOpen
  \bibfield  {author} {\bibinfo {author} {\bibfnamefont {M.}~\bibnamefont
  {Merkel}}\ and\ \bibinfo {author} {\bibfnamefont {M.~L.}\ \bibnamefont
  {Manning}},\ }\bibfield  {title} {\bibinfo {title} {A geometrically
  controlled rigidity transition in a model for confluent 3d tissues},\
  }\href@noop {} {\bibfield  {journal} {\bibinfo  {journal} {New Journal of
  Physics}\ }\textbf {\bibinfo {volume} {20}},\ \bibinfo {pages} {022002}
  (\bibinfo {year} {2018})}\BibitemShut {NoStop}%
\end{thebibliography}

%

\end{document}